\documentclass[letter]{aa}
\usepackage[varg]{txfonts}
\usepackage{graphicx}
\usepackage{lscape}
\usepackage[breaklinks]{hyperref}
\usepackage{siunitx}
\usepackage{orcidlink}

\renewcommand{\linenumbers}{}

\newcommand{\ur}[1]{\,\mathrm{#1}} 

\newcommand{\norm}[1]{\lVert #1\rVert }

\defcitealias{Paper1}{Paper~I}

\definecolor{cobalt}{rgb}{0.06, 0.2, 0.65}
\hypersetup{
  colorlinks,
  citecolor=cobalt,
  linkcolor=cobalt,
  urlcolor=cobalt,
}

\usepackage{listings}
\definecolor{dkgreen}{rgb}{0,0.6,0}
\definecolor{gray}{rgb}{0.5,0.5,0.5}
\definecolor{mauve}{rgb}{0.58,0,0.82}
\begin{document}

\title{Solar twins in Gaia DR3 GSP-Spec}
\subtitle{II. Age distribution and its implications for the Sun's migration}

\author{
    Takuji Tsujimoto\inst{\ref{inst:NAOJ}}\orcidlink{0000-0002-9397-3658}
    \and 
    Daisuke Taniguchi\inst{\ref{inst:TMU},\ref{inst:NAOJ}}\fnmsep\thanks{\scriptsize The Tokyo Center For Excellence Project, Tokyo Metropolitan University}\orcidlink{0000-0002-2861-4069}
    \and 
    Alejandra Recio-Blanco\inst{\ref{inst:OCA}}\orcidlink{0000-0002-6550-7377}
    \and 
    Pedro A. Palicio\inst{\ref{inst:OCA}}\orcidlink{0000-0002-7432-8709}
    \and 
    Patrick de Laverny \inst{\ref{inst:OCA}}\orcidlink{0000-0002-2817-4104}
}

\institute{
    National Astronomical Observatory of Japan, 2-21-1 Osawa, Mitaka, Tokyo 181-8588, Japan \\
    \email{taku.tsujimoto@nao.ac.jp}\label{inst:NAOJ}
    \and 
    Department of Physics, Tokyo Metropolitan University, 1-1 Minami-Osawa, Hachioji, Tokyo 192-0397, Japan \\
    \email{d.taniguchi.astro@gmail.com}\label{inst:TMU}
    \and 
    Universit\'e C\^ote d'Azur, Observatoire de la C\^ote d'Azur, CNRS, Laboratoire Lagrange, Bd de l'Observatoire, CS 34229, 06304 Nice cedex 4, France\label{inst:OCA}
}

\date{Received ?? ; accepted ??}

\abstract{Solar twins are among the most powerful tracers of Galactic disk evolution owing to their unique property of sharing nearly solar metallicities ($\text{[Fe/H]}\approx 0$) while spanning a wide range of ages. To grasp solar twins as relics of Galaxy evolution, individual twins must be tagged with ages. A sufficiently large and well-characterized stellar sample then allows us to construct an age distribution that encodes the star formation history beyond our local region, modulated by the efficiency of radial migration of stars. Based on our catalog of $6{,}594$ high-quality local (${\lesssim }300\ur{pc}$) solar twins from the \textit{Gaia} Data Release 3 spectroscopic (\textit{GSP-Spec}) catalog, we derived their age distribution after carefully deconvolving the selection function. We find two distinct features:~a narrow peak around ${\sim }2\ur{Gyr}$ and a broad bump extending over ${\sim }4\text{--}6\ur{Gyr}$. First, we argue that the former corresponds to a relatively recent burst of star formation that occurred in the disk, including at least a local region within a few kiloparsecs of the Sun, which is in good agreement with previous results deduced from independent works. On the other hand, the older bump, closely associated with the Sun's birth epoch, is intriguing since this finding challenges the predicted presence of a corotation barrier built by the Galactic bar, which is thought to prevent stars born inside $R_{\mathrm{GC}}\approx 6\ur{kpc}$ from reaching the solar neighborhood. We propose that the large number of local twins with ages between $4$ and $6\ur{Gyr}$ provides compelling evidence that the Sun's long-distance ($\geq 3\ur{kpc}$) migration is shared by many inner disk stars. This, in turn, suggests a possible link with the epoch of bar formation, which may have triggered enhanced star formation in the inner disk and induced efficient radial migration. }

\keywords{
    Galaxy: disk --
    Stars: solar-type --
    Galaxy: evolution --
    Stars: statistics --
    Galaxy: kinematics and dynamics
}

\titlerunning{Solar twins in Gaia DR3 GSP-Spec. II}

\maketitle 

\section{Introduction}\label{Sec:Intro}

Clear answers to the question of how the Galaxy disk forms and evolves through cosmic time continue to elude us. Recent studies have deepened the mystery via findings of the Galaxy's complex structure and history, rendering our simplistic initial picture outdated \citep[e.g.][]{2015ApJ...808..132H, 2022Natur.603..599X}. As an established scheme, the Galactic disk is composed of two distinguishable populations in terms of chemistry and/or kinematics, the so-called thick and thin disks \citep[e.g.,][for a review]{2016ARA&A..54..529B}; however, it is still debatable how each disk was formed and whether their origins are connected. Of the two disks, the thin disk lies at the forefront of Galaxy formation, not least because most stars are in the thin disk today. The discovery of very metal-poor ($\text{[Fe/H]}<-2.5$) thin disk stars \citep[e.g.,][]{2020MNRAS.497L...7S} including a RR Lyrae star (${>}10\ur{Gyr}$ old) by \citet{2022ApJ...925...10M} suggests that the thin disk may have extended back to the early days of the Galaxy. This finding is closely tied to the debated origin of these stars \citep[e.g.,][]{2021MNRAS.500.3750S, 2022MNRAS.514..689B, 2024A&A...683A.136B, 2024A&A...690A..26L}, highlighting the complex history of disk (for simplicity, we hereafter refer to the thin disk as the disk). 

More than two decades ago, it had already been suggested that star formation in the disk exhibits a complex history, with several bursts having taken place in the past \citep{2000A&A...358..869R}. Since then, this view has become more compelling, aided by the growing body of observational data thanks, in part, to large surveys \citep[e.g.,][]{2019A&A...624L...1M, 2022Natur.603..599X}, as well as the advancement in techniques for determining the star formation history \citep[e.g.,][]{2005ARA&A..43..387G}. In particular, the \textit{Gaia} mission has made a breakthrough in this field by providing precise measurements of stellar parallaxes, which have dramatically improved the quality of the color-magnitude diagram \citep{2018A&A...616A..10G}. This update leads to a more accurate reconstruction of the star formation history, revealing clear evidence of episodic (bursting) star formation in the disk throughout much of cosmic time, including the events around ${\sim }2$ and ${\sim }6\ur{Gyr}$ ago \citep{2020NatAs...4..965R}. This finding has promoted the argument that the bursting events could have been triggered by episodic encounters between the Sagittarius (Sgr) dwarf galaxy and the Galaxy. It supports a dynamic perspective in which the complex history of the Galaxy disk is shaped by external influences, possibly beginning with a major impact from Gaia-Enceladus \citep{2018Natur.563...85H, 2018MNRAS.478..611B}. 

In addition, the emerging paradigm of Galactic dynamics highlights the crucial role of internal interactions between stars and the Galactic structure within the disk in shaping its evolution:~stars migrate radially across the disk through resonant scattering with spiral arms and the central bar over their lifetimes \citep[e.g.,][]{2002MNRAS.336..785S, 2008ApJ...684L..79R, 2009MNRAS.396..203S, 2010ApJ...722..112M}. This so-called radial migration of stars predicts that the birthplaces of a considerable number of disk stars differ from their current positions, and that the stars in the solar vicinity are a mixture of populations born at a wide range of Galactocentric distances ($R_{\mathrm{GC}}$) across the disk. Accordingly, the renewed view of Galactic dynamics suggests that the locally determined star formation history is, in fact, the outcome of a complex assembly of Galactic internal and external events that have occurred across nearly the entire disk over cosmic time. 

In this context, it is of particular importance to classify local stars according to their approximate birth radii within the disk. This can in fact be achieved through chemical tagging combined with stellar age estimates, based on our conventional understanding that the chemical evolution of the disk varies with $R_{\mathrm{GC}}$, resulting in a distinct age-metallicity relation for each radius. This view is observationally evidenced by current radial abundance gradients showing a higher metallicity at smaller $R_{\mathrm{GC}}$ values. Such a negative abundance gradient is naturally explained in the context of galaxy formation, with the inner region forming faster and becoming more metal-rich than the outer region -- a process known as the inside-out scenario \citep{2001ApJ...554.1044C}. In other words, the rate at which metallicity reaches a given value, such as the solar metallicity, increases with decreasing $R_{\mathrm{GC}}$. 

There exist stars nearly identical to the Sun, so-called solar twins, which exhibit stellar atmospheric characteristics, including metallicity, quite similar to the solar values. \citet{2018ApJ...865...68B} demonstrate that local $79$ solar twins exhibit a tight correlation between the [X/Fe] ratios (where X ranges from C to Dy) and stellar ages widely distributed from $0$ to $9\ur{Gyr}$. Furthermore, this correlation is shown to be interpreted on the basis of the age--$R_{\mathrm{GC}}$ connection within the framework of Galactic chemical evolution \citep{2021ApJ...920L..32T}. Then, suppose that a statistically sufficient number of age-tagged solar twins were in hand: we would be able to explore the Galaxy disk's past through a novel methodology, one that could allow us to probe a vast region, including the unexplored innermost disk \citep{2024A&A...691A.298P}. The best way to get as large as possible a catalog of good solar twin candidates is to dig into the catalogs published by the large spectroscopic surveys. The largest one in number is the homogeneous General Stellar Parametrizer from Spectroscopy (\textit{GSP-Spec}) catalog collected from space by the ESA/\textit{Gaia} mission and published as part of the third \textit{Gaia} data release \citep[DR3,][]{2023A&A...674A..29R}. \textit{GSP-Spec} relies on the analysis of ${\sim }5.6$ million \textit{Gaia}/Radial Velocity Spectrometer stellar spectra. 

We emphasize that the number of local solar twins as a function of stellar age represents a specific fraction of stars formed at their birth radii that were able to migrate to the solar vicinity, governed by the efficiency of radial migration. In this context, particular attention should be paid to a potential dynamical mechanism that could hinder the outward migration of stars from the inner disk:~the Galactic bar creates a corotation barrier due to constraints imposed by the conservation of Jacobi energy \citep{2008gady.book.....B}. This mechanism would predict a significantly reduced number of old solar twins in the solar neighborhood. A representative case is provided by numerical simulations investigating the Sun's migration from its presumed birth radius of $R_{\mathrm{GC}}\approx 5\ur{kpc}$ \citep{2020ApJ...904..137T, 2024ApJ...976L..29B}. These studies suggest that only ${\sim }1\%$ or fewer of stars sharing the Sun's birth radius are able to reach the solar vicinity within the Sun's lifetime of $4.6\ur{Gyr}$. Indeed, the age distribution of solar twins may serve as a critical test of this hypothesis.

\vspace{-0.4cm}
\section{Derivation of intrinsic age distribution}\label{sec:DataSet}

From the \textit{Gaia} DR3 GSP-Spec catalog, \citet[hereafter \citetalias{Paper1}]{Paper1} built a catalog of $6{,}594$ solar twins with a well-characterized selection function. Accurate ages of individual twins were estimated with the isochrone-projection method, in which the observed $T_{\mathrm{eff}}$, [M/H], and either $\log g$, $M_{G}$, or $M_{K_{\mathrm{s}}}$\footnote{In this Letter, we present results based on the ages inferred from $M_{K_{\mathrm{s}}}$, since this quantity is only minimally affected by selection effects. The results based on the ages from $\log g$ and $M_{G}$ are provided in Appendix~\ref{app:figures} for reference. } were compared against PARSEC isochrones version 1.2S~\citep{2012MNRAS.427..127B,2015MNRAS.452.1068C}. The resultant catalog, which covers a volume of ${\sim }20\text{--}300\ur{pc}$, is two orders of magnitude larger than the previous ones that observed individual stars one by one (see Table~2 in \citetalias{Paper1}). \citetalias{Paper1} also constructed a mock catalog to characterize the observational selection effects (see Sect.~4.3 in \citetalias{Paper1} for details). 

\begin{figure}
\centering 
\includegraphics[width=9cm]{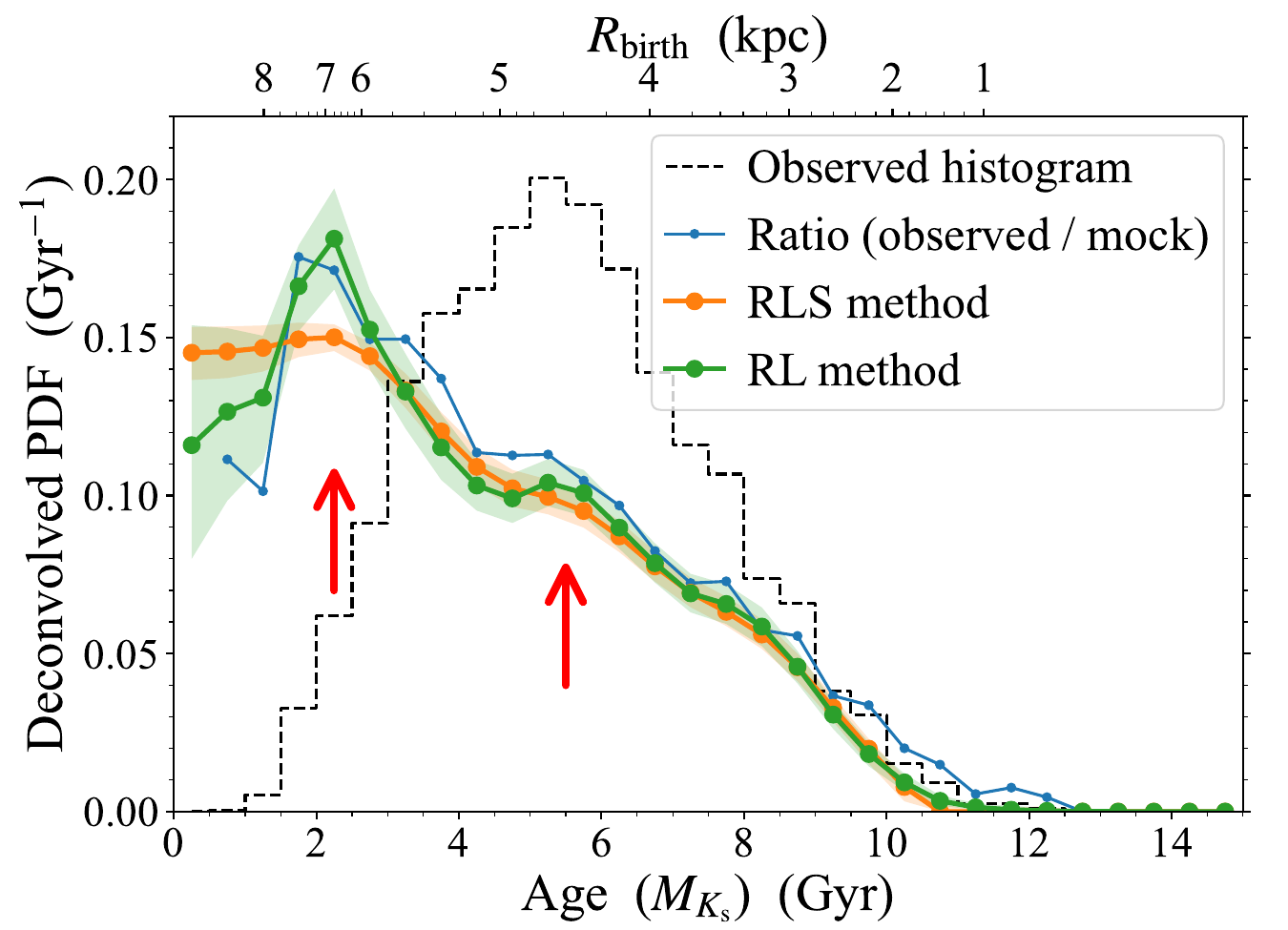}
\vspace{-0.4cm}
\caption{(Deconvolved) probability density function (PDF) of the ages of solar twins determined with $M_{K_{\mathrm{s}}}$. The dotted black line shows the normalized age histogram of our solar twin catalog (same as the blue histogram in Fig.~\ref{fig:AgeHist0_K0}). The blue line shows the normalized ratio between the observed and mock age histograms, as a rough estimate of the intrinsic age distribution. Orange and green lines represent the deconvolved, intrinsic age distributions obtained using the two different (RLS and RL) methods. Shaded regions indicate the statistical uncertainty in the deconvolved PDFs, estimated from $10^{4}$ Monte Carlo realizations assuming Poisson statistics for the original age histogram (see Appendix~\ref{app:StatBias} for a validation of this uncertainty estimate). The upper axis provides reference $R_{\mathrm{birth}}$ values, based on the $\text{[Fe/H]}=0\ur{dex}$ slice of the age--$R_{\mathrm{birth}}$--[Fe/H] relation in \citet{2023MNRAS.525.2208R}. See Fig.~\ref{fig:DeconvHist_logg_G0} for results using $\log g$ and those using $M_{G}$. }
\label{fig:DeconvHist_K0}
\end{figure}

The purpose of this Letter is to derive the ``intrinsic'' age distribution of migrated stars, which is defined as the product of the star formation rate at their birth radii and the migration efficiency. For this purpose, we first obtained a rough estimate of the intrinsic age distribution of solar twins from the ratio of the observed (blue) to mock (orange) age histograms shown in Fig.~\ref{fig:AgeHist0_K0}. In addition, we deconvolved the selection function from the observed solar twin age histogram using two independent approaches:~regularized least-squares (RLS) inversion and Richardson-Lucy (RL) iterative deconvolution methods (see Appendix~\ref{app:deconvolve} for details). The results obtained with the three methods are shown in Fig.~\ref{fig:DeconvHist_K0} by blue, orange, and green lines. 

We find that the overall shapes of the intrinsic age distributions are qualitatively similar:~each shows at least two prominent peaks (or bumps, identified by red arrows in Fig.~\ref{fig:DeconvHist_K0}) around $2\ur{Gyr}$ and $4\text{--}6\ur{Gyr}$, with the exception that the latter bump is not present in the distribution derived using the RLS method. In the following discussion, we focus on these features, although we note that an additional bump may also be present around $8\ur{Gyr}$. Furthermore, the presence of solar twins as old as ${\sim }10\ur{Gyr}$ suggests that the formation of the disk may have begun more than at least $10\ur{Gyr}$ ago. 

\vspace{-0.4cm}
\section{Bimodal age distribution}\label{sec:agedist}

The deduced age distribution exhibits a peak corresponding to young solar twins and a secondary bump comprising relatively old twins with ages of $4\text{--}6\ur{Gyr}$. In this study, because the age uncertainty is as large as $2\ur{Gyr}$, it is difficult to resolve the age distribution within the most recent $2\ur{Gyr}$. However, we can at least conclude that the clear increase from ${\sim }3\ur{Gyr}$ ago toward the present points to the existence of a dominant population of young solar twins at these ages. Several studies have reported peaks corresponding to relatively young local stars, such as at ${\sim }2.5\ur{Gyr}$ ago \citep{2019A&A...624L...1M, 2022MNRAS.510.4669S, 2025A&A...694A.120C}, as well as $1.9$ and $1\ur{Gyr}$ ago \citep{2020NatAs...4..965R}. Considering the age uncertainties in our study, our results are broadly consistent with these findings over a relatively large volume of a few kiloparsecs, corresponding to the possible travel distances of $2\ur{Gyr}$-old stars \citep[approximately $1\text{--}1.5\ur{kpc}$;][]{2020ApJ...904..137T}, similar to the case of \citet{2020NatAs...4..965R}. Note that our solar twin sample cannot probe very recent star formation, in contrast to previous studies based on samples that include OBA-type stars. Accordingly, we argue that a burst of star formation took place in the Galactic disk, encompassing at least the local region around $2\ur{Gyr}$ ago. It is worth reiterating here that its possible trigger is an interaction with the Sgr dwarf galaxy, which deserves further attention. 

More intriguingly, we identify a bump feature formed by solar twins with ages of ${\sim }4\text{--}6\ur{Gyr}$. Evidence for enhanced star formation at these epochs has also been reported in previous studies \citep[e.g.,][]{2020NatAs...4..965R, 2024A&A...687A.168G, 2025A&A...697A.128D}\footnote{In \citet{2024A&A...687A.168G}, the age distribution is presented as a function of stellar metallicity. According to their results, the bump around $4\text{--}6\ur{Gyr}$ is composed mainly of stars with metallicities lower than the solar value, contrary to our solar twin sample. }. However, our result differs in one important respect:~in our sample, the bump stars consist exclusively of stars that migrated from the inner disk, following the solar twin's age--$R_{\mathrm{GC}}$ relation, whereas in other studies where no metallicity constraint is imposed on sample selection, they represent a mixture of stars formed both in situ and in other regions of the disk. This broad coincidence between samples with and without in situ star formation can be reconciled with a framework that the locally inferred star formation history may have been substantially shaped by the effects of radial migration \citep[e.g.,][]{2025PASJ...77..916B}. The question then arises of where these $4\text{--}6\ur{Gyr}$-old stars migrated to the solar neighborhood from. The approximate birthplaces of such stars can be inferred from the studies so far, including the one illustrated in Fig.~\ref{fig:DeconvHist_K0}. It is worth emphasizing here that solar twins with ages ranging from $4.5$ to $7\ur{Gyr}$ exhibit elemental abundance patterns that match the Sun's almost perfectly \citep{2021ApJ...920L..32T}, suggesting that the bump stars likely share a very similar birth environment to that of the Sun. 

Although some uncertainty remains in the estimated birth radius of the Sun, most studies broadly converge on a value of $R_{\mathrm{GC}}<6\ur{kpc}$, with specific estimates including ${\sim }4.5\ur{kpc}$ \citep{2024MNRAS.535..392L}, ${\sim }5\ur{kpc}$ \citep{2023MNRAS.525.2208R, 2023MNRAS.526.6088B}, and ${\sim }6\ur{kpc}$ \citep{2023MNRAS.523.2126P}, based on placing the Sun within the framework of Galactic chemical evolution. However, from the standpoint of Galactic dynamics, migration from such an inner region may not be sufficiently probable. This argument stems from the presence of the Galactic bar, which generates a potential barrier that makes it difficult for stars to migrate outward across the corotation radius \citep[CR;][]{2002MNRAS.336..785S, 2007MNRAS.379.1155C, 2013A&A...553A.102D, 2019MNRAS.488.3324F, 2020A&A...638A.144K}, located around $R_{\mathrm{GC}}\sim 6\ur{kpc}$ \citep[e.g.,][]{2017MNRAS.465.1621P, 2019MNRAS.488.4552S} under the slow-long-bar scenario \citep[e.g.,][]{2018MNRAS.477.3945H, 2018MNRAS.478.1231P}. Nevertheless, the bar length and pattern speed remain open issues, with some studies favoring a short-fast-bar scenario \citep{2000AJ....119..800D, 2018MNRAS.474...95H}. In addition, the Galactic bar is thought to be decelerating \citep{2025ApJ...983L..10Z, 2025MNRAS.542.1331D}, with an estimated present-day slowing rate of $\Omega_{\mathrm{p}} = -4.5 \pm 1.4\ur{\si{km.s^{-1}.kpc^{-1}}}$ \citep{2021MNRAS.500.4710C}. As a consequence of the mentioned potential barrier, the likelihood of such large-scale outward migration is expected to be quite low and has  been estimated to be on the order of ${\sim }1\%$ or less \citep{2020ApJ...904..137T, 2024ApJ...976L..29B}. This implies that only a negligible number of solar twins with ages close to that of the Sun (${\approx }4\text{--}6\ur{Gyr}$) should exist, given that their birthplaces lie within the corotation radius. Yet, contrary to this expectation, our derived age distribution shows no dip in this age range; instead, it exhibits a clear bump feature. 

This naturally raises the question of what mechanism is responsible for producing such a feature. A plausible explanation is that a particular event or process temporarily enhanced the efficiency of radial migration in the inner disk. This scenario is reminiscent of the epoch of Galactic bar formation, which could have induced a coupling effect, simultaneously enhancing both radial migration efficiency \citep[e.g.,][]{1995A&A...301..649F} and star formation \citep[e.g.,][]{1992MNRAS.259..328A}. In particular, the combined effect of the bar and the spiral-arm structure can induce stronger migration around the bar's CR \citep[][but see also \cite{2003AJ....125..785Q}]{2010ApJ...722..112M}. If this theoretical prediction applies, the bump feature in our age distribution may mark the onset of bar formation around $6\text{--}7\ur{Gyr}$ ago, with the forming bar remaining dynamically active for a few gigayears thereafter. As a result, the overall epoch associated with bar formation activity can be identified as the period ${\sim }4\text{--}7\ur{Gyr}$ ago, corresponding approximately to the full extent of the bump feature. This inferred bar age is somewhat younger than the commonly cited value of ${\sim }8\ur{Gyr}$ \citep[e.g.,][]{2019MNRAS.490.4740B}; however, a recent study based on the ages of super-metal-rich stars suggests a much younger bar age of $3\text{--}4\ur{Gyr}$ \citep{2024A&A...681L...8N}, partially overlapping with our estimate. Furthermore, a bar formation epoch around $4\ur{Gyr}$ ago has also been proposed from a perspective similar to ours \citep{2026A&A...705A..92R}. Our argument is supported by the recent finding of an inner stellar ring at the bar ends, which is composed of stars with metallicities approximately close to solar and ages overlapping with those of bump twins \citep{2022A&A...659A..80W}. 

Finally, we discuss the bimodality in the age distribution purely from the perspective of chemical evolution. If the thin disk formed as a secondary component after the thick disk, it is expected to have originated from super-metal-rich gas ($\text{[Fe/H]}>0$), after which [Fe/H] would have decreased over time due to dilution by infalling metal-poor gas. This reverse evolution phase would have ended relatively quickly, followed by normal chemical evolution. This scenario naturally predicts a bimodality in the ages of solar twins, yielding both young (formed within the last few gigayears) and old populations. However, model calculations suggest that the old twins should be as old as ${\sim }10\ur{Gyr}$ and fewer in number than what is observed \citep[e.g.,][]{2023A&A...678A..61P}.

\vspace{-0.4cm}
\section{Orbital evolution}

\begin{figure}
\centering 
\includegraphics[width=9cm]{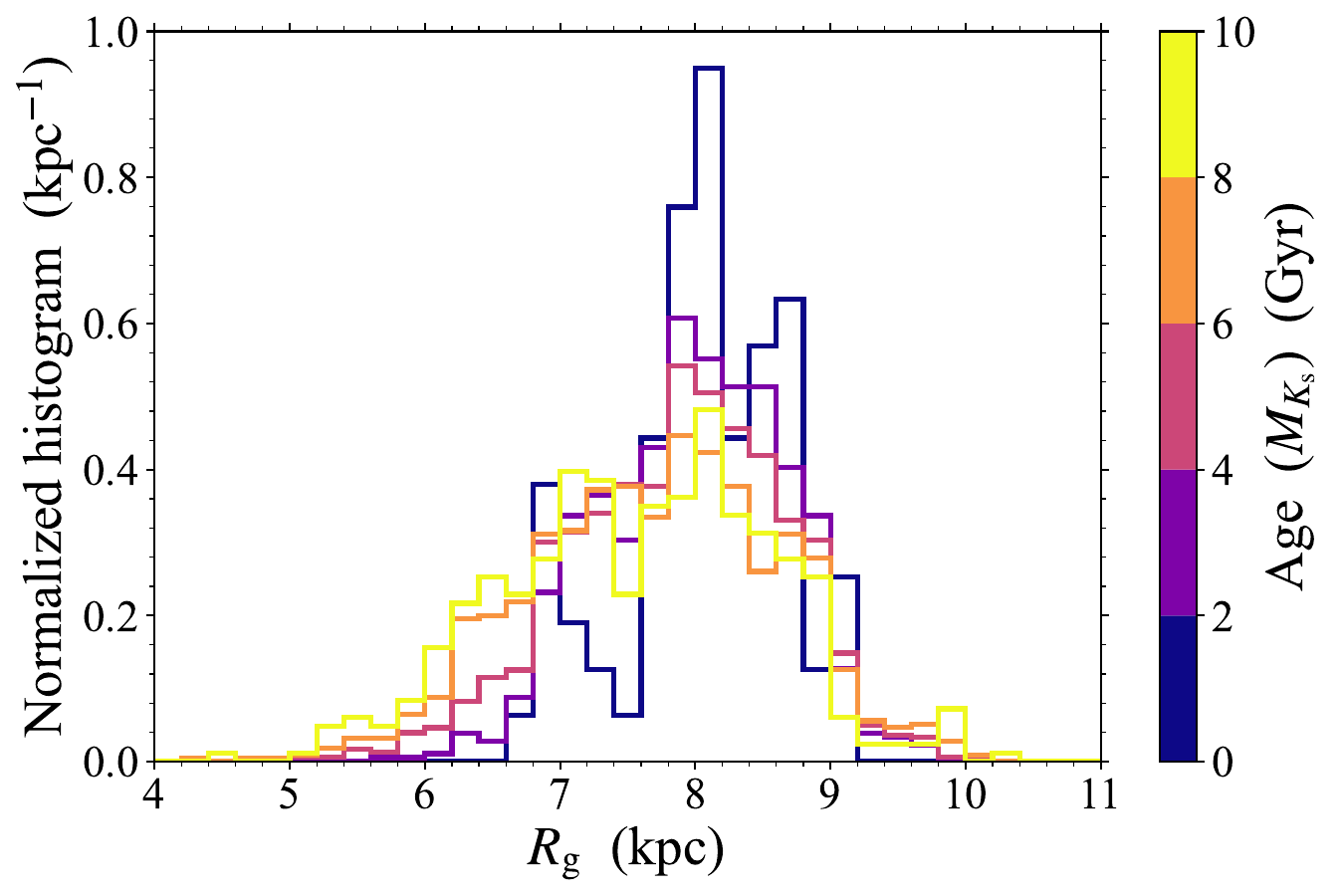}
\vspace{-0.4cm}
\caption{Normalized histograms of the guiding radii, $R_{\mathrm{g}}$, for solar twins, color-coded by their ages determined with $M_{K_{\mathrm{s}}}$. Histograms for other orbital parameters are shown in Fig.~\ref{fig:OrbitParams_K0_ecc_zmax}. }
\label{fig:OrbitParams_K0_Rg}
\end{figure}

The arguments presented in Sect.~\ref{sec:agedist} are framed within the context of radial migration theory, in which older solar twins are interpreted as having migrated from inner regions closer to the Galactic center, under the assumption that their orbits have been altered primarily through a dynamically cold process: so-called ``churning.'' To test this hypothesis, we examined the orbital property of solar twins as a function of age. Figure~\ref{fig:OrbitParams_K0_Rg} presents a histogram, divided into age bins, of the guiding radii ($R_{\mathrm{g}}$). No distinct differences are found among the age-tagged distributions, except for a slightly broader distribution for older twins, which is within the expected range for the churning process. A similar feature is also seen in the distributions of eccentricities, $e$, and maximum vertical distances from the disk, $Z_{\mathrm{max}}$, as shown in Fig.~\ref{fig:OrbitParams_K0_ecc_zmax}. We therefore conclude that the orbital evolution of solar twins has been governed primarily by diffusion in angular momentum rather than by radial heating (``blurring''), as assumed. If the blurring process dominated instead, the peaks in the $R_{\mathrm{g}}$ distributions for older twins would shift to smaller values than $8\ur{kpc}$, reflecting their inner birth radii.

\vspace{-0.4cm}
\section{Conclusions}\label{sec:conclusion}

From our catalog of $6{,}594$ local solar twins, whose ages were carefully determined in our separate paper~\citepalias{Paper1}, we derived their age distribution. This provides synchronized information on both the star formation history of the disk within the solar radius and the radial migration of the corresponding stars. We find a bimodal feature, with a prominent peak around ${\sim }2\ur{Gyr}$ ago and a broad bump extending over ${\sim }4\text{--}6\ur{Gyr}$ ago. The former is consistent with previous studies suggesting that a burst of star formation was triggered relatively recently in the Galactic disk encompassing the solar neighborhood. Our result strengthens the argument that interactions with the Sgr dwarf galaxy during pericentric passages may have induced bursts of star formation in the disk. 

In contrast, the latter may indicate the important role of the Galactic bar, which can enhance both star formation and the efficiency of radial migration during its formation epoch. This interpretation points to a relatively young age for the bar, estimated to be ${\sim }4\text{--}7\ur{Gyr}$, corresponding to a period of heightened activity within the forming bar. This phase of strong bar activity likely overlapped with the formation of the Solar System in the inner disk. In this context, the Sun would have naturally migrated as one member of a large group of co-migrating stars.

\bibliographystyle{aa} 
\bibliography{ref}

@ARTICLE{Paper1,
       author = {{Taniguchi}, Daisuke and {de Laverny}, Patrick and {Recio-Blanco}, Alejandra and {Tsujimoto}, Takuji and {Palicio}, Pedro~A.},
        title = "{Solar twins in Gaia DR3 GSP-Spec: I. Building a large catalog of Solar twins with ages}",
      journal = {\aap},
         year = 2026,
       volume = {XXX},
          eid = {AXXX},
        pages = {AXXX},
          doi = {10.1051/0004-6361/XXXXXXXXX},
archivePrefix = {arXiv},
       eprint = {XXXX.XXXXX},
 primaryClass = {astro-ph.SR},
       adsurl = {https://ui.adsabs.harvard.edu/abs/2026A&A...XXXA.XXXT}
}

@ARTICLE{2015ApJ...808..132H,
       author = {{Hayden}, Michael R. and {Bovy}, Jo and {Holtzman}, Jon A. and {Nidever}, David L. and {Bird}, Jonathan C. and {Weinberg}, David H. and {Andrews}, Brett H. and {Majewski}, Steven R. and {Allende Prieto}, Carlos and {Anders}, Friedrich and {Beers}, Timothy C. and {Bizyaev}, Dmitry and {Chiappini}, Cristina and {Cunha}, Katia and {Frinchaboy}, Peter and {Garc{\'\i}a-Her{\'n}andez}, D.~A. and {Garc{\'\i}a P{\'e}rez}, Ana E. and {Girardi}, L{\'e}o and {Harding}, Paul and {Hearty}, Fred R. and {Johnson}, Jennifer A. and {M{\'e}sz{\'a}ros}, Szabolcs and {Minchev}, Ivan and {O'Connell}, Robert and {Pan}, Kaike and {Robin}, Annie C. and {Schiavon}, Ricardo P. and {Schneider}, Donald P. and {Schultheis}, Mathias and {Shetrone}, Matthew and {Skrutskie}, Michael and {Steinmetz}, Matthias and {Smith}, Verne and {Wilson}, John C. and {Zamora}, Olga and {Zasowski}, Gail},
        title = "{Chemical Cartography with APOGEE: Metallicity Distribution Functions and the Chemical Structure of the Milky Way Disk}",
      journal = {\apj},
         year = 2015,
        month = aug,
       volume = {808},
       number = {2},
          eid = {132},
        pages = {132},
          doi = {10.1088/0004-637X/808/2/132},
archivePrefix = {arXiv},
       eprint = {1503.02110},
 primaryClass = {astro-ph.GA},
       adsurl = {https://ui.adsabs.harvard.edu/abs/2015ApJ...808..132H}
}

@ARTICLE{2022Natur.603..599X,
       author = {{Xiang}, Maosheng and {Rix}, Hans-Walter},
        title = "{A time-resolved picture of our Milky Way's early formation history}",
      journal = {\nat},
         year = 2022,
        month = mar,
       volume = {603},
       number = {7902},
        pages = {599-603},
          doi = {10.1038/s41586-022-04496-5},
archivePrefix = {arXiv},
       eprint = {2203.12110},
 primaryClass = {astro-ph.GA},
       adsurl = {https://ui.adsabs.harvard.edu/abs/2022Natur.603..599X}
}

@ARTICLE{2016ARA&A..54..529B,
       author = {{Bland-Hawthorn}, Joss and {Gerhard}, Ortwin},
        title = "{The Galaxy in Context: Structural, Kinematic, and Integrated Properties}",
      journal = {\araa},
         year = 2016,
        month = sep,
       volume = {54},
        pages = {529-596},
          doi = {10.1146/annurev-astro-081915-023441},
archivePrefix = {arXiv},
       eprint = {1602.07702},
 primaryClass = {astro-ph.GA},
       adsurl = {https://ui.adsabs.harvard.edu/abs/2016ARA&A..54..529B}
}

@ARTICLE{2020MNRAS.497L...7S,
       author = {{Sestito}, Federico and {Martin}, Nicolas F. and {Starkenburg}, Else and {Arentsen}, Anke and {Ibata}, Rodrigo A. and {Longeard}, Nicolas and {Kielty}, Collin and {Youakim}, Kristopher and {Venn}, Kim A. and {Aguado}, David S. and {Carlberg}, Raymond G. and {Gonz{\'a}lez Hern{\'a}ndez}, Jonay I. and {Hill}, Vanessa and {Jablonka}, Pascale and {Kordopatis}, Georges and {Malhan}, Khyati and {Navarro}, Julio F. and {S{\'a}nchez-Janssen}, Rub{\'e}n and {Thomas}, Guillame and {Tolstoy}, Eline and {Wilson}, Thomas G. and {Palicio}, Pedro A. and {Bialek}, Spencer and {Garcia-Dias}, Rafael and {Lucchesi}, Romain and {North}, Pierre and {Osorio}, Yeisson and {Patrick}, Lee R. and {Peralta de Arriba}, Luis},
        title = "{The Pristine survey - X. A large population of low-metallicity stars permeates the Galactic disc}",
      journal = {\mnras},
         year = 2020,
        month = sep,
       volume = {497},
       number = {1},
        pages = {L7-L12},
          doi = {10.1093/mnrasl/slaa022},
archivePrefix = {arXiv},
       eprint = {1911.08491},
 primaryClass = {astro-ph.GA},
       adsurl = {https://ui.adsabs.harvard.edu/abs/2020MNRAS.497L...7S}
}

@ARTICLE{2022ApJ...925...10M,
       author = {{Matsunaga}, Noriyuki and {Itane}, Akinori and {Hattori}, Kohei and {Crestani}, Juliana and {Braga}, Vittorio and {Bono}, Giuseppe and {Taniguchi}, Daisuke and {Baba}, Junichi and {Maehara}, Hiroyuki and {Ukita}, Nobuharu and {Sakamoto}, Tsuyoshi and {Kobayashi}, Naoto and {Aoki}, Tsutomu and {Soyano}, Takao and {Tarusawa}, Ken'ichi and {Sarugaku}, Yuki and {Mito}, Hiroyuki and {Sako}, Shigeyuki and {Doi}, Mamoru and {Nakada}, Yoshikazu and {Izumi}, Natsuko and {Ita}, Yoshifusa and {Onozato}, Hiroki and {Jian}, Mingjie and {Kondo}, Sohei and {Hamano}, Satoshi and {Yasui}, Chikako and {Tsujimoto}, Takuji and {Otsubo}, Shogo and {Ikeda}, Yuji and {Kawakita}, Hideyo},
        title = "{A Very Metal-poor RR Lyrae Star with a Disk Orbit Found in the Solar Neighborhood}",
      journal = {\apj},
         year = 2022,
        month = jan,
       volume = {925},
       number = {1},
          eid = {10},
        pages = {10},
          doi = {10.3847/1538-4357/ac3483},
archivePrefix = {arXiv},
       eprint = {2201.05402},
 primaryClass = {astro-ph.SR},
       adsurl = {https://ui.adsabs.harvard.edu/abs/2022ApJ...925...10M}
}

@ARTICLE{2021MNRAS.500.3750S,
       author = {{Sestito}, Federico and {Buck}, Tobias and {Starkenburg}, Else and {Martin}, Nicolas F. and {Navarro}, Julio F. and {Venn}, Kim A. and {Obreja}, Aura and {Jablonka}, Pascale and {Macci{\`o}}, Andrea V.},
        title = "{Exploring the origin of low-metallicity stars in Milky-Way-like galaxies with the NIHAO-UHD simulations}",
      journal = {\mnras},
         year = 2021,
        month = jan,
       volume = {500},
       number = {3},
        pages = {3750-3762},
          doi = {10.1093/mnras/staa3479},
archivePrefix = {arXiv},
       eprint = {2009.14207},
 primaryClass = {astro-ph.GA},
       adsurl = {https://ui.adsabs.harvard.edu/abs/2021MNRAS.500.3750S}
}

@ARTICLE{2022MNRAS.514..689B,
       author = {{Belokurov}, Vasily and {Kravtsov}, Andrey},
        title = "{From dawn till disc: Milky Way's turbulent youth revealed by the APOGEE+Gaia data}",
      journal = {\mnras},
         year = 2022,
        month = jul,
       volume = {514},
       number = {1},
        pages = {689-714},
          doi = {10.1093/mnras/stac1267},
archivePrefix = {arXiv},
       eprint = {2203.04980},
 primaryClass = {astro-ph.GA},
       adsurl = {https://ui.adsabs.harvard.edu/abs/2022MNRAS.514..689B}
}

@ARTICLE{2024A&A...683A.136B,
       author = {{Bellazzini}, M. and {Massari}, D. and {Ceccarelli}, E. and {Mucciarelli}, A. and {Bragaglia}, A. and {Riello}, M. and {De Angeli}, F. and {Montegriffo}, P.},
        title = "{Metal-poor stars with disc-like orbits. Possible traces of the Galactic disc at very early epochs}",
      journal = {\aap},
         year = 2024,
        month = mar,
       volume = {683},
          eid = {A136},
        pages = {A136},
          doi = {10.1051/0004-6361/202348106},
archivePrefix = {arXiv},
       eprint = {2312.02356},
 primaryClass = {astro-ph.GA},
       adsurl = {https://ui.adsabs.harvard.edu/abs/2024A&A...683A.136B}
}

@ARTICLE{2024A&A...690A..26L,
       author = {{Li}, Chengdong and {Yuan}, Zhen and {Monari}, Giacomo and {Martin}, Nicolas F. and {Siebert}, Arnaud and {Famaey}, Benoit and {Chiba}, Rimpei and {Kordopatis}, Georges and {Ibata}, Rodrigo A. and {Hill}, Vanessa},
        title = "{Exploring the impact of a decelerating bar on transforming bulge orbits into disc-like orbits}",
      journal = {\aap},
         year = 2024,
        month = oct,
       volume = {690},
          eid = {A26},
        pages = {A26},
          doi = {10.1051/0004-6361/202449742},
archivePrefix = {arXiv},
       eprint = {2311.15270},
 primaryClass = {astro-ph.GA},
       adsurl = {https://ui.adsabs.harvard.edu/abs/2024A&A...690A..26L}
}

@ARTICLE{2000A&A...358..869R,
       author = {{Rocha-Pinto}, H.~J. and {Scalo}, J. and {Maciel}, W.~J. and {Flynn}, C.},
        title = "{Chemical enrichment and star formation in the Milky Way disk. II. Star formation history}",
      journal = {\aap},
         year = 2000,
        month = jun,
       volume = {358},
        pages = {869-885},
          doi = {10.48550/arXiv.astro-ph/0001383},
archivePrefix = {arXiv},
       eprint = {astro-ph/0001383},
 primaryClass = {astro-ph},
       adsurl = {https://ui.adsabs.harvard.edu/abs/2000A&A...358..869R}
}

@ARTICLE{2019A&A...624L...1M,
       author = {{Mor}, R. and {Robin}, A.~C. and {Figueras}, F. and {Roca-F{\`a}brega}, S. and {Luri}, X.},
        title = "{Gaia DR2 reveals a star formation burst in the disc 2-3 Gyr ago}",
      journal = {\aap},
         year = 2019,
        month = apr,
       volume = {624},
          eid = {L1},
        pages = {L1},
          doi = {10.1051/0004-6361/201935105},
archivePrefix = {arXiv},
       eprint = {1901.07564},
 primaryClass = {astro-ph.GA},
       adsurl = {https://ui.adsabs.harvard.edu/abs/2019A&A...624L...1M}
}

@ARTICLE{2005ARA&A..43..387G,
       author = {{Gallart}, C. and {Zoccali}, M. and {Aparicio}, A.},
        title = "{The Adequacy of Stellar Evolution Models for the Interpretation of the Color-Magnitude Diagrams of Resolved Stellar Populations}",
      journal = {\araa},
         year = 2005,
        month = sep,
       volume = {43},
       number = {1},
        pages = {387-434},
          doi = {10.1146/annurev.astro.43.072103.150608},
       adsurl = {https://ui.adsabs.harvard.edu/abs/2005ARA&A..43..387G}
}

@ARTICLE{2018A&A...616A..10G,
       author = {{Gaia Collaboration} and {Babusiaux}, C. and {van Leeuwen}, F. and {Barstow}, M.~A. and {Jordi}, C. and {Vallenari}, A. and {Bossini}, D. and {Bressan}, A. and {Cantat-Gaudin}, T. and {van Leeuwen}, M. and {Brown}, A.~G.~A. and {Prusti}, T. and {de Bruijne}, J.~H.~J. and {Bailer-Jones}, C.~A.~L. and {Biermann}, M. and {Evans}, D.~W. and {Eyer}, L. and {Jansen}, F. and {Klioner}, S.~A. and {Lammers}, U. and {Lindegren}, L. and {Luri}, X. and {Mignard}, F. and {Panem}, C. and {Pourbaix}, D. and {Randich}, S. and {Sartoretti}, P. and {Siddiqui}, H.~I. and {Soubiran}, C. and {Walton}, N.~A. and {Arenou}, F. and {Bastian}, U. and {Cropper}, M. and {Drimmel}, R. and {Katz}, D. and {Lattanzi}, M.~G. and {Bakker}, J. and {Cacciari}, C. and {Casta{\~n}eda}, J. and {Chaoul}, L. and {Cheek}, N. and {De Angeli}, F. and {Fabricius}, C. and {Guerra}, R. and {Holl}, B. and {Masana}, E. and {Messineo}, R. and {Mowlavi}, N. and {Nienartowicz}, K. and {Panuzzo}, P. and {Portell}, J. and {Riello}, M. and {Seabroke}, G.~M. and {Tanga}, P. and {Th{\'e}venin}, F. and {Gracia-Abril}, G. and {Comoretto}, G. and {Garcia-Reinaldos}, M. and {Teyssier}, D. and {Altmann}, M. and {Andrae}, R. and {Audard}, M. and {Bellas-Velidis}, I. and {Benson}, K. and {Berthier}, J. and {Blomme}, R. and {Burgess}, P. and {Busso}, G. and {Carry}, B. and {Cellino}, A. and {Clementini}, G. and {Clotet}, M. and {Creevey}, O. and {Davidson}, M. and {De Ridder}, J. and {Delchambre}, L. and {Dell'Oro}, A. and {Ducourant}, C. and {Fern{\'a}ndez-Hern{\'a}ndez}, J. and {Fouesneau}, M. and {Fr{\'e}mat}, Y. and {Galluccio}, L. and {Garc{\'\i}a-Torres}, M. and {Gonz{\'a}lez-N{\'u}{\~n}ez}, J. and {Gonz{\'a}lez-Vidal}, J.~J. and {Gosset}, E. and {Guy}, L.~P. and {Halbwachs}, J. -L. and {Hambly}, N.~C. and {Harrison}, D.~L. and {Hern{\'a}ndez}, J. and {Hestroffer}, D. and {Hodgkin}, S.~T. and {Hutton}, A. and {Jasniewicz}, G. and {Jean-Antoine-Piccolo}, A. and {Jordan}, S. and {Korn}, A.~J. and {Krone-Martins}, A. and {Lanzafame}, A.~C. and {Lebzelter}, T. and {L{\"o}ffler}, W. and {Manteiga}, M. and {Marrese}, P.~M. and {Mart{\'\i}n-Fleitas}, J.~M. and {Moitinho}, A. and {Mora}, A. and {Muinonen}, K. and {Osinde}, J. and {Pancino}, E. and {Pauwels}, T. and {Petit}, J. -M. and {Recio-Blanco}, A. and {Richards}, P.~J. and {Rimoldini}, L. and {Robin}, A.~C. and {Sarro}, L.~M. and {Siopis}, C. and {Smith}, M. and {Sozzetti}, A. and {S{\"u}veges}, M. and {Torra}, J. and {van Reeven}, W. and {Abbas}, U. and {Abreu Aramburu}, A. and {Accart}, S. and {Aerts}, C. and {Altavilla}, G. and {{\'A}lvarez}, M.~A. and {Alvarez}, R. and {Alves}, J. and {Anderson}, R.~I. and {Andrei}, A.~H. and {Anglada Varela}, E. and {Antiche}, E. and {Antoja}, T. and {Arcay}, B. and {Astraatmadja}, T.~L. and {Bach}, N. and {Baker}, S.~G. and {Balaguer-N{\'u}{\~n}ez}, L. and {Balm}, P. and {Barache}, C. and {Barata}, C. and {Barbato}, D. and {Barblan}, F. and {Barklem}, P.~S. and {Barrado}, D. and {Barros}, M. and {Bartholom{\'e} Mu{\~n}oz}, L. and {Bassilana}, J. -L. and {Becciani}, U. and {Bellazzini}, M. and {Berihuete}, A. and {Bertone}, S. and {Bianchi}, L. and {Bienaym{\'e}}, O. and {Blanco-Cuaresma}, S. and {Boch}, T. and {Boeche}, C. and {Bombrun}, A. and {Borrachero}, R. and {Bouquillon}, S. and {Bourda}, G. and {Bragaglia}, A. and {Bramante}, L. and {Breddels}, M.~A. and {Brouillet}, N. and {Br{\"u}semeister}, T. and {Brugaletta}, E. and {Bucciarelli}, B. and {Burlacu}, A. and {Busonero}, D. and {Butkevich}, A.~G. and {Buzzi}, R. and {Caffau}, E. and {Cancelliere}, R. and {Cannizzaro}, G. and {Carballo}, R. and {Carlucci}, T. and {Carrasco}, J.~M. and {Casamiquela}, L. and {Castellani}, M. and {Castro-Ginard}, A. and {Charlot}, P. and {Chemin}, L. and {Chiavassa}, A. and {Cocozza}, G. and {Costigan}, G. and {Cowell}, S. and {Crifo}, F. and {Crosta}, M. and {Crowley}, C. and {Cuypers}, J. and {Dafonte}, C. and {Damerdji}, Y. and {Dapergolas}, A. and {David}, P. and {David}, M. and {de Laverny}, P.},
        title = "{Gaia Data Release 2. Observational Hertzsprung-Russell diagrams}",
      journal = {\aap},
         year = 2018,
        month = aug,
       volume = {616},
          eid = {A10},
        pages = {A10},
          doi = {10.1051/0004-6361/201832843},
archivePrefix = {arXiv},
       eprint = {1804.09378},
 primaryClass = {astro-ph.SR},
       adsurl = {https://ui.adsabs.harvard.edu/abs/2018A&A...616A..10G}
}

@ARTICLE{2020NatAs...4..965R,
       author = {{Ruiz-Lara}, Tom{\'a}s and {Gallart}, Carme and {Bernard}, Edouard J. and {Cassisi}, Santi},
        title = "{The recurrent impact of the Sagittarius dwarf on the star formation history of the Milky Way}",
      journal = {Nature Astronomy},
         year = 2020,
        month = may,
       volume = {4},
        pages = {965-973},
          doi = {10.1038/s41550-020-1097-0},
archivePrefix = {arXiv},
       eprint = {2003.12577},
 primaryClass = {astro-ph.GA},
       adsurl = {https://ui.adsabs.harvard.edu/abs/2020NatAs...4..965R}
}

@ARTICLE{2018Natur.563...85H,
       author = {{Helmi}, Amina and {Babusiaux}, Carine and {Koppelman}, Helmer H. and {Massari}, Davide and {Veljanoski}, Jovan and {Brown}, Anthony G.~A.},
        title = "{The merger that led to the formation of the Milky Way's inner stellar halo and thick disk}",
      journal = {\nat},
         year = 2018,
        month = oct,
       volume = {563},
       number = {7729},
        pages = {85-88},
          doi = {10.1038/s41586-018-0625-x},
archivePrefix = {arXiv},
       eprint = {1806.06038},
 primaryClass = {astro-ph.GA},
       adsurl = {https://ui.adsabs.harvard.edu/abs/2018Natur.563...85H}
}

@ARTICLE{2018MNRAS.478..611B,
       author = {{Belokurov}, V. and {Erkal}, D. and {Evans}, N.~W. and {Koposov}, S.~E. and {Deason}, A.~J.},
        title = "{Co-formation of the disc and the stellar halo}",
      journal = {\mnras},
         year = 2018,
        month = jul,
       volume = {478},
       number = {1},
        pages = {611-619},
          doi = {10.1093/mnras/sty982},
archivePrefix = {arXiv},
       eprint = {1802.03414},
 primaryClass = {astro-ph.GA},
       adsurl = {https://ui.adsabs.harvard.edu/abs/2018MNRAS.478..611B}
}

@ARTICLE{2002MNRAS.336..785S,
       author = {{Sellwood}, J.~A. and {Binney}, J.~J.},
        title = "{Radial mixing in galactic discs}",
      journal = {\mnras},
         year = 2002,
        month = nov,
       volume = {336},
       number = {3},
        pages = {785-796},
          doi = {10.1046/j.1365-8711.2002.05806.x},
archivePrefix = {arXiv},
       eprint = {astro-ph/0203510},
 primaryClass = {astro-ph},
       adsurl = {https://ui.adsabs.harvard.edu/abs/2002MNRAS.336..785S}
}

@ARTICLE{2008ApJ...684L..79R,
       author = {{Ro{\v{s}}kar}, Rok and {Debattista}, Victor P. and {Quinn}, Thomas R. and {Stinson}, Gregory S. and {Wadsley}, James},
        title = "{Riding the Spiral Waves: Implications of Stellar Migration for the Properties of Galactic Disks}",
      journal = {\apjl},
         year = 2008,
        month = sep,
       volume = {684},
       number = {2},
        pages = {L79},
          doi = {10.1086/592231},
archivePrefix = {arXiv},
       eprint = {0808.0206},
 primaryClass = {astro-ph},
       adsurl = {https://ui.adsabs.harvard.edu/abs/2008ApJ...684L..79R}
}

@ARTICLE{2009MNRAS.396..203S,
       author = {{Sch{\"o}nrich}, Ralph and {Binney}, James},
        title = "{Chemical evolution with radial mixing}",
      journal = {\mnras},
         year = 2009,
        month = jun,
       volume = {396},
       number = {1},
        pages = {203-222},
          doi = {10.1111/j.1365-2966.2009.14750.x},
archivePrefix = {arXiv},
       eprint = {0809.3006},
 primaryClass = {astro-ph},
       adsurl = {https://ui.adsabs.harvard.edu/abs/2009MNRAS.396..203S}
}

@ARTICLE{2010ApJ...722..112M,
       author = {{Minchev}, I. and {Famaey}, B.},
        title = "{A New Mechanism for Radial Migration in Galactic Disks: Spiral-Bar Resonance Overlap}",
      journal = {\apj},
         year = 2010,
        month = oct,
       volume = {722},
       number = {1},
        pages = {112-121},
          doi = {10.1088/0004-637X/722/1/112},
archivePrefix = {arXiv},
       eprint = {0911.1794},
 primaryClass = {astro-ph.GA},
       adsurl = {https://ui.adsabs.harvard.edu/abs/2010ApJ...722..112M}
}

@ARTICLE{2001ApJ...554.1044C,
       author = {{Chiappini}, Cristina and {Matteucci}, Francesca and {Romano}, Donatella},
        title = "{Abundance Gradients and the Formation of the Milky Way}",
      journal = {\apj},
         year = 2001,
        month = jun,
       volume = {554},
       number = {2},
        pages = {1044-1058},
          doi = {10.1086/321427},
archivePrefix = {arXiv},
       eprint = {astro-ph/0102134},
 primaryClass = {astro-ph},
       adsurl = {https://ui.adsabs.harvard.edu/abs/2001ApJ...554.1044C}
}

@ARTICLE{2018ApJ...865...68B,
       author = {{Bedell}, Megan and {Bean}, Jacob L. and {Mel{\'e}ndez}, Jorge and {Spina}, Lorenzo and {Ram{\'\i}rez}, Ivan and {Asplund}, Martin and {Alves-Brito}, Alan and {dos Santos}, Leonardo and {Dreizler}, Stefan and {Yong}, David and {Monroe}, TalaWanda and {Casagrande}, Luca},
        title = "{The Chemical Homogeneity of Sun-like Stars in the Solar Neighborhood}",
      journal = {\apj},
         year = 2018,
        month = sep,
       volume = {865},
       number = {1},
          eid = {68},
        pages = {68},
          doi = {10.3847/1538-4357/aad908},
archivePrefix = {arXiv},
       eprint = {1802.02576},
 primaryClass = {astro-ph.SR},
       adsurl = {https://ui.adsabs.harvard.edu/abs/2018ApJ...865...68B}
}

@ARTICLE{2021ApJ...920L..32T,
       author = {{Tsujimoto}, Takuji},
        title = "{Two Sites of r-process Production Assessed on the Basis of the Age-tagged Abundances of Solar Twins}",
      journal = {\apjl},
         year = 2021,
        month = oct,
       volume = {920},
       number = {2},
          eid = {L32},
        pages = {L32},
          doi = {10.3847/2041-8213/ac2c75},
archivePrefix = {arXiv},
       eprint = {2110.02261},
 primaryClass = {astro-ph.GA},
       adsurl = {https://ui.adsabs.harvard.edu/abs/2021ApJ...920L..32T}
}

@ARTICLE{2024A&A...691A.298P,
       author = {{Plotnikova}, A. and {Spina}, L. and {Ratcliffe}, B. and {Casali}, G. and {Carraro}, G.},
        title = "{The chemical evolution of the Milky Way thin disk using solar twins}",
      journal = {\aap},
         year = 2024,
        month = nov,
       volume = {691},
          eid = {A298},
        pages = {A298},
          doi = {10.1051/0004-6361/202451167},
archivePrefix = {arXiv},
       eprint = {2411.03067},
 primaryClass = {astro-ph.GA},
       adsurl = {https://ui.adsabs.harvard.edu/abs/2024A&A...691A.298P}
}

@ARTICLE{2023A&A...674A..29R,
       author = {{Recio-Blanco}, A. and {de Laverny}, P. and {Palicio}, P.~A. and {Kordopatis}, G. and {{\'A}lvarez}, M.~A. and {Schultheis}, M. and {Contursi}, G. and {Zhao}, H. and {Torralba Elipe}, G. and {Ordenovic}, C. and {Manteiga}, M. and {Dafonte}, C. and {Oreshina-Slezak}, I. and {Bijaoui}, A. and {Fr{\'e}mat}, Y. and {Seabroke}, G. and {Pailler}, F. and {Spitoni}, E. and {Poggio}, E. and {Creevey}, O.~L. and {Abreu Aramburu}, A. and {Accart}, S. and {Andrae}, R. and {Bailer-Jones}, C.~A.~L. and {Bellas-Velidis}, I. and {Brouillet}, N. and {Brugaletta}, E. and {Burlacu}, A. and {Carballo}, R. and {Casamiquela}, L. and {Chiavassa}, A. and {Cooper}, W.~J. and {Dapergolas}, A. and {Delchambre}, L. and {Dharmawardena}, T.~E. and {Drimmel}, R. and {Edvardsson}, B. and {Fouesneau}, M. and {Garabato}, D. and {Garc{\'\i}a-Lario}, P. and {Garc{\'\i}a-Torres}, M. and {Gavel}, A. and {Gomez}, A. and {Gonz{\'a}lez-Santamar{\'\i}a}, I. and {Hatzidimitriou}, D. and {Heiter}, U. and {Jean-Antoine Piccolo}, A. and {Kontizas}, M. and {Korn}, A.~J. and {Lanzafame}, A.~C. and {Lebreton}, Y. and {Le Fustec}, Y. and {Licata}, E.~L. and {Lindstr{\o}m}, H.~E.~P. and {Livanou}, E. and {Lobel}, A. and {Lorca}, A. and {Magdaleno Romeo}, A. and {Marocco}, F. and {Marshall}, D.~J. and {Mary}, N. and {Nicolas}, C. and {Pallas-Quintela}, L. and {Panem}, C. and {Pichon}, B. and {Riclet}, F. and {Robin}, C. and {Rybizki}, J. and {Santove{\~n}a}, R. and {Silvelo}, A. and {Smart}, R.~L. and {Sarro}, L.~M. and {Sordo}, R. and {Soubiran}, C. and {S{\"u}veges}, M. and {Ulla}, A. and {Vallenari}, A. and {Zorec}, J. and {Utrilla}, E. and {Bakker}, J.},
        title = "{Gaia Data Release 3. Analysis of RVS spectra using the General Stellar Parametriser from spectroscopy}",
      journal = {\aap},
         year = 2023,
        month = jun,
       volume = {674},
          eid = {A29},
        pages = {A29},
          doi = {10.1051/0004-6361/202243750},
archivePrefix = {arXiv},
       eprint = {2206.05541},
 primaryClass = {astro-ph.GA},
       adsurl = {https://ui.adsabs.harvard.edu/abs/2023A&A...674A..29R}
}

@BOOK{2008gady.book.....B,
       author = {{Binney}, James and {Tremaine}, Scott},
        title = "{Galactic Dynamics: Second Edition}",
         year = 2008,
       adsurl = {https://ui.adsabs.harvard.edu/abs/2008gady.book.....B},
    publisher = {Princeton Univ. Press},
      address = {Princeton}
}

@ARTICLE{2020ApJ...904..137T,
       author = {{Tsujimoto}, Takuji and {Baba}, Junichi},
        title = "{Remarkable Migration of the Solar System from the Innermost Galactic Disk; a Wander, a Wobble, and a Climate Catastrophe on the Earth}",
      journal = {\apj},
         year = 2020,
        month = dec,
       volume = {904},
       number = {2},
          eid = {137},
        pages = {137},
          doi = {10.3847/1538-4357/abc00a},
archivePrefix = {arXiv},
       eprint = {2010.05962},
 primaryClass = {astro-ph.GA},
       adsurl = {https://ui.adsabs.harvard.edu/abs/2020ApJ...904..137T}
}

@ARTICLE{2024ApJ...976L..29B,
       author = {{Baba}, Junichi and {Tsujimoto}, Takuji and {Saitoh}, Takayuki R.},
        title = "{Solar System Migration Points to a Renewed Concept: Galactic Habitable Orbits}",
      journal = {\apjl},
         year = 2024,
        month = dec,
       volume = {976},
       number = {2},
          eid = {L29},
        pages = {L29},
          doi = {10.3847/2041-8213/ad9260},
archivePrefix = {arXiv},
       eprint = {2412.02963},
 primaryClass = {astro-ph.GA},
       adsurl = {https://ui.adsabs.harvard.edu/abs/2024ApJ...976L..29B}
}

@ARTICLE{2012MNRAS.427..127B,
       author = {{Bressan}, Alessandro and {Marigo}, Paola and {Girardi}, L{\'e}o. and {Salasnich}, Bernardo and {Dal Cero}, Claudia and {Rubele}, Stefano and {Nanni}, Ambra},
        title = "{PARSEC: stellar tracks and isochrones with the PAdova and TRieste Stellar Evolution Code}",
      journal = {\mnras},
         year = 2012,
        month = nov,
       volume = {427},
       number = {1},
        pages = {127-145},
          doi = {10.1111/j.1365-2966.2012.21948.x},
archivePrefix = {arXiv},
       eprint = {1208.4498},
 primaryClass = {astro-ph.SR},
       adsurl = {https://ui.adsabs.harvard.edu/abs/2012MNRAS.427..127B}
}

@ARTICLE{2015MNRAS.452.1068C,
       author = {{Chen}, Yang and {Bressan}, Alessandro and {Girardi}, L{\'e}o and {Marigo}, Paola and {Kong}, Xu and {Lanza}, Antonio},
        title = "{PARSEC evolutionary tracks of massive stars up to 350 M$_{{\ensuremath{\odot}}}$ at metallicities 0.0001 {\ensuremath{\leq}} Z {\ensuremath{\leq}} 0.04}",
      journal = {\mnras},
         year = 2015,
        month = sep,
       volume = {452},
       number = {1},
        pages = {1068-1080},
          doi = {10.1093/mnras/stv1281},
archivePrefix = {arXiv},
       eprint = {1506.01681},
 primaryClass = {astro-ph.SR},
       adsurl = {https://ui.adsabs.harvard.edu/abs/2015MNRAS.452.1068C}
}

@ARTICLE{2023MNRAS.525.2208R,
       author = {{Ratcliffe}, Bridget and {Minchev}, Ivan and {Anders}, Friedrich and {Khoperskov}, Sergey and {Guiglion}, Guillaume and {Buck}, Tobias and {Cunha}, Katia and {Queiroz}, Anna and {Nitschelm}, Christian and {Meszaros}, Szabolcs and {Steinmetz}, Matthias and {de Jong}, Roelof S. and {Nepal}, Samir and {Lane}, Richard R. and {Sobeck}, Jennifer},
        title = "{Unveiling the time evolution of chemical abundances across the Milky Way disc with APOGEE}",
      journal = {\mnras},
         year = 2023,
        month = oct,
       volume = {525},
       number = {2},
        pages = {2208-2228},
          doi = {10.1093/mnras/stad1573},
archivePrefix = {arXiv},
       eprint = {2305.13378},
 primaryClass = {astro-ph.GA},
       adsurl = {https://ui.adsabs.harvard.edu/abs/2023MNRAS.525.2208R}
}

@ARTICLE{2022MNRAS.510.4669S,
       author = {{Sahlholdt}, Christian L. and {Feltzing}, Sofia and {Feuillet}, Diane K.},
        title = "{Characterizing epochs of star formation across the Milky Way disc using age-metallicity distributions of GALAH stars}",
      journal = {\mnras},
         year = 2022,
        month = mar,
       volume = {510},
       number = {4},
        pages = {4669-4688},
          doi = {10.1093/mnras/stab3681},
archivePrefix = {arXiv},
       eprint = {2112.08218},
 primaryClass = {astro-ph.GA},
       adsurl = {https://ui.adsabs.harvard.edu/abs/2022MNRAS.510.4669S}
}

@ARTICLE{2025A&A...694A.120C,
       author = {{Chen}, Tianxiang and {Prantzos}, Nikos},
        title = "{Recent star formation episodes in the Galaxy: Impact on its chemical properties and the evolution of its abundance gradient}",
      journal = {\aap},
         year = 2025,
        month = feb,
       volume = {694},
          eid = {A120},
        pages = {A120},
          doi = {10.1051/0004-6361/202452552},
archivePrefix = {arXiv},
       eprint = {2501.03342},
 primaryClass = {astro-ph.GA},
       adsurl = {https://ui.adsabs.harvard.edu/abs/2025A&A...694A.120C}
}

@ARTICLE{2024A&A...687A.168G,
       author = {{Gallart}, Carme and {Surot}, Francisco and {Cassisi}, Santi and {Fern{\'a}ndez-Alvar}, Emma and {Mirabal}, David and {Rivero}, Alicia and {Ruiz-Lara}, Tom{\'a}s and {Santos-Torres}, Judith and {Aznar-Menargues}, Guillem and {Battaglia}, Giuseppina and {Queiroz}, Anna B. and {Monelli}, Matteo and {Vasiliev}, Eugene and {Chiappini}, Cristina and {Helmi}, Amina and {Hill}, Vanessa and {Massari}, Davide and {Thomas}, Guillaume F.},
        title = "{Chronology of our Galaxy from Gaia colour-magnitude diagram fitting (ChronoGal). I. The formation and evolution of the thin disc from the Gaia Catalogue of Nearby Stars}",
      journal = {\aap},
         year = 2024,
        month = jul,
       volume = {687},
          eid = {A168},
        pages = {A168},
          doi = {10.1051/0004-6361/202349078},
archivePrefix = {arXiv},
       eprint = {2402.09399},
 primaryClass = {astro-ph.GA},
       adsurl = {https://ui.adsabs.harvard.edu/abs/2024A&A...687A.168G}
}

@ARTICLE{2025A&A...697A.128D,
       author = {{del Alc{\'a}zar-Juli{\`a}}, M. and {Figueras}, F. and {Robin}, A.~C. and {Bienaym{\'e}}, O. and {Anders}, F.},
        title = "{Joint inference of the Milky Way's star formation history and initial mass function from Gaia all-sky G < 13 data}",
      journal = {\aap},
         year = 2025,
        month = may,
       volume = {697},
          eid = {A128},
        pages = {A128},
          doi = {10.1051/0004-6361/202453606},
archivePrefix = {arXiv},
       eprint = {2501.17236},
 primaryClass = {astro-ph.GA},
       adsurl = {https://ui.adsabs.harvard.edu/abs/2025A&A...697A.128D}
}

@ARTICLE{2025PASJ...77..916B,
       author = {{Baba}, Junichi},
        title = "{Influence of bar formation on star formation segregation and stellar migration: Implications for variations in the age distribution of Milky Way disk stars}",
      journal = {\pasj},
         year = 2025,
        month = aug,
       volume = {77},
       number = {4},
        pages = {916-923},
          doi = {10.1093/pasj/psaf062},
archivePrefix = {arXiv},
       eprint = {2505.16528},
 primaryClass = {astro-ph.GA},
       adsurl = {https://ui.adsabs.harvard.edu/abs/2025PASJ...77..916B}
}

@ARTICLE{2024MNRAS.535..392L,
       author = {{Lu}, Yuxi (Lucy) and {Minchev}, Ivan and {Buck}, Tobias and {Khoperskov}, Sergey and {Steinmetz}, Matthias and {Libeskind}, Noam and {Cescutti}, Gabriele and {Freeman}, Ken C. and {Ratcliffe}, Bridget},
        title = "{There is no place like home - finding birth radii of stars in the Milky Way}",
      journal = {\mnras},
         year = 2024,
        month = nov,
       volume = {535},
       number = {1},
        pages = {392-405},
          doi = {10.1093/mnras/stae2364},
archivePrefix = {arXiv},
       eprint = {2212.04515},
 primaryClass = {astro-ph.GA},
       adsurl = {https://ui.adsabs.harvard.edu/abs/2024MNRAS.535..392L}
}

@ARTICLE{2023MNRAS.526.6088B,
       author = {{Baba}, Junichi and {Saitoh}, Takayuki R. and {Tsujimoto}, Takuji},
        title = "{Exploring the Sun's birth radius and the distribution of planet building blocks in the Milky Way galaxy: a multizone Galactic chemical evolution approach}",
      journal = {\mnras},
         year = 2023,
        month = dec,
       volume = {526},
       number = {4},
        pages = {6088-6102},
          doi = {10.1093/mnras/stad3188},
archivePrefix = {arXiv},
       eprint = {2310.10335},
 primaryClass = {astro-ph.GA},
       adsurl = {https://ui.adsabs.harvard.edu/abs/2023MNRAS.526.6088B}
}

@ARTICLE{2023MNRAS.523.2126P,
       author = {{Prantzos}, Nikos and {Abia}, Carlos and {Chen}, Tianxiang and {de Laverny}, Patrick and {Recio-Blanco}, Alejandra and {Athanassoula}, E. and {Roberti}, Lorenzo and {Vescovi}, Diego and {Limongi}, Marco and {Chieffi}, Alessandro and {Cristallo}, Sergio},
        title = "{On the origin of the Galactic thin and thick discs, their abundance gradients and the diagnostic potential of their abundance ratios}",
      journal = {\mnras},
         year = 2023,
        month = aug,
       volume = {523},
       number = {2},
        pages = {2126-2145},
          doi = {10.1093/mnras/stad1551},
archivePrefix = {arXiv},
       eprint = {2305.13431},
 primaryClass = {astro-ph.GA},
       adsurl = {https://ui.adsabs.harvard.edu/abs/2023MNRAS.523.2126P}
}

@ARTICLE{2007MNRAS.379.1155C,
       author = {{Ceverino}, D. and {Klypin}, A.},
        title = "{Resonances in barred galaxies}",
      journal = {\mnras},
         year = 2007,
        month = aug,
       volume = {379},
       number = {3},
        pages = {1155-1168},
          doi = {10.1111/j.1365-2966.2007.12001.x},
archivePrefix = {arXiv},
       eprint = {astro-ph/0703544},
 primaryClass = {astro-ph},
       adsurl = {https://ui.adsabs.harvard.edu/abs/2007MNRAS.379.1155C}
}

@ARTICLE{2013A&A...553A.102D,
       author = {{Di Matteo}, P. and {Haywood}, M. and {Combes}, F. and {Semelin}, B. and {Snaith}, O.~N.},
        title = "{Signatures of radial migration in barred galaxies: Azimuthal variations in the metallicity distribution of old stars}",
      journal = {\aap},
         year = 2013,
        month = may,
       volume = {553},
          eid = {A102},
        pages = {A102},
          doi = {10.1051/0004-6361/201220539},
archivePrefix = {arXiv},
       eprint = {1301.2545},
 primaryClass = {astro-ph.GA},
       adsurl = {https://ui.adsabs.harvard.edu/abs/2013A&A...553A.102D}
}

@ARTICLE{2019MNRAS.488.3324F,
       author = {{Fragkoudi}, F. and {Katz}, D. and {Trick}, W. and {White}, S.~D.~M. and {Di Matteo}, P. and {Sormani}, M.~C. and {Khoperskov}, S. and {Haywood}, M. and {Hall{\'e}}, A. and {G{\'o}mez}, A.},
        title = "{On the ridges, undulations, and streams in Gaia DR2: linking the topography of phase space to the orbital structure of an N-body bar}",
      journal = {\mnras},
         year = 2019,
        month = sep,
       volume = {488},
       number = {3},
        pages = {3324-3339},
          doi = {10.1093/mnras/stz1875},
archivePrefix = {arXiv},
       eprint = {1901.07568},
 primaryClass = {astro-ph.GA},
       adsurl = {https://ui.adsabs.harvard.edu/abs/2019MNRAS.488.3324F}
}

@ARTICLE{2020A&A...638A.144K,
       author = {{Khoperskov}, S. and {Di Matteo}, P. and {Haywood}, M. and {G{\'o}mez}, A. and {Snaith}, O.~N.},
        title = "{Escapees from the bar resonances. Presence of low-eccentricity metal-rich stars at the solar vicinity}",
      journal = {\aap},
         year = 2020,
        month = jun,
       volume = {638},
          eid = {A144},
        pages = {A144},
          doi = {10.1051/0004-6361/201937188},
archivePrefix = {arXiv},
       eprint = {1911.12424},
 primaryClass = {astro-ph.GA},
       adsurl = {https://ui.adsabs.harvard.edu/abs/2020A&A...638A.144K}
}

@ARTICLE{2017MNRAS.465.1621P,
       author = {{Portail}, Matthieu and {Gerhard}, Ortwin and {Wegg}, Christopher and {Ness}, Melissa},
        title = "{Dynamical modelling of the galactic bulge and bar: the Milky Way's pattern speed, stellar and dark matter mass distribution}",
      journal = {\mnras},
         year = 2017,
        month = feb,
       volume = {465},
       number = {2},
        pages = {1621-1644},
          doi = {10.1093/mnras/stw2819},
archivePrefix = {arXiv},
       eprint = {1608.07954},
 primaryClass = {astro-ph.GA},
       adsurl = {https://ui.adsabs.harvard.edu/abs/2017MNRAS.465.1621P}
}

@ARTICLE{2019MNRAS.488.4552S,
       author = {{Sanders}, Jason L. and {Smith}, Leigh and {Evans}, N. Wyn},
        title = "{The pattern speed of the Milky Way bar from transverse velocities}",
      journal = {\mnras},
         year = 2019,
        month = oct,
       volume = {488},
       number = {4},
        pages = {4552-4564},
          doi = {10.1093/mnras/stz1827},
archivePrefix = {arXiv},
       eprint = {1903.02009},
 primaryClass = {astro-ph.GA},
       adsurl = {https://ui.adsabs.harvard.edu/abs/2019MNRAS.488.4552S}
}

@ARTICLE{2018MNRAS.477.3945H,
       author = {{Hunt}, Jason A.~S. and {Bovy}, Jo},
        title = "{The 4:1 outer Lindblad resonance of a long-slow bar as an explanation for the Hercules stream}",
      journal = {\mnras},
         year = 2018,
        month = jul,
       volume = {477},
       number = {3},
        pages = {3945-3953},
          doi = {10.1093/mnras/sty921},
archivePrefix = {arXiv},
       eprint = {1803.02358},
 primaryClass = {astro-ph.GA},
       adsurl = {https://ui.adsabs.harvard.edu/abs/2018MNRAS.477.3945H}
}

@ARTICLE{2018MNRAS.478.1231P,
       author = {{Palicio}, Pedro A. and {Martinez-Valpuesta}, Inma and {Allende Prieto}, Carlos and {Dalla Vecchia}, Claudio and {Zamora}, Olga and {Zasowski}, Gail and {Fernandez-Trincado}, J.~G. and {Masters}, Karen L. and {Garc{\'\i}a-Hern{\'a}ndez}, D.~A. and {Roman-Lopes}, Alexandre},
        title = "{Signatures of the Galactic bar on stellar kinematics unveiled by APOGEE}",
      journal = {\mnras},
         year = 2018,
        month = jul,
       volume = {478},
       number = {1},
        pages = {1231-1243},
          doi = {10.1093/mnras/sty1156},
archivePrefix = {arXiv},
       eprint = {1805.04347},
 primaryClass = {astro-ph.GA},
       adsurl = {https://ui.adsabs.harvard.edu/abs/2018MNRAS.478.1231P}
}

@ARTICLE{2000AJ....119..800D,
       author = {{Dehnen}, Walter},
        title = "{The Effect of the Outer Lindblad Resonance of the Galactic Bar on the Local Stellar Velocity Distribution}",
      journal = {\aj},
         year = 2000,
        month = feb,
       volume = {119},
       number = {2},
        pages = {800-812},
          doi = {10.1086/301226},
archivePrefix = {arXiv},
       eprint = {astro-ph/9911161},
 primaryClass = {astro-ph},
       adsurl = {https://ui.adsabs.harvard.edu/abs/2000AJ....119..800D}
}

@ARTICLE{2018MNRAS.474...95H,
       author = {{Hunt}, Jason A.~S. and {Bovy}, Jo and {P{\'e}rez-Villegas}, Angeles and {Holtzman}, Jon A. and {Sobeck}, Jennifer and {Chojnowski}, Drew and {Santana}, Felipe A. and {Palicio}, Pedro A. and {Wegg}, Christopher and {Gerhard}, Ortwin and {Almeida}, Andr{\'e}s and {Bizyaev}, Dmitry and {Fernandez-Trincado}, Jose G. and {Lane}, Richard R. and {Longa-Pe{\~n}a}, Pen{\'e}lope and {Majewski}, Steven R. and {Pan}, Kaike and {Roman-Lopes}, Alexandre},
        title = "{The Hercules stream as seen by APOGEE-2 South}",
      journal = {\mnras},
         year = 2018,
        month = feb,
       volume = {474},
       number = {1},
        pages = {95-101},
          doi = {10.1093/mnras/stx2777},
archivePrefix = {arXiv},
       eprint = {1709.02807},
 primaryClass = {astro-ph.GA},
       adsurl = {https://ui.adsabs.harvard.edu/abs/2018MNRAS.474...95H}
}

@ARTICLE{2025ApJ...983L..10Z,
       author = {{Zhang}, HanYuan and {Belokurov}, Vasily and {Evans}, N. Wyn and {Sanders}, Jason L. and {Lu}, Yuxi(Lucy) and {Cao}, Chengye and {Myeong}, GyuChul and {Dillamore}, Adam M. and {Kane}, Sarah G. and {Li}, Zhao-Yu},
        title = "{Observational Constraints of Radial Migration in the Galactic Disk Driven by the Slowing Bar}",
      journal = {\apjl},
         year = 2025,
        month = apr,
       volume = {983},
       number = {1},
          eid = {L10},
        pages = {L10},
          doi = {10.3847/2041-8213/adc261},
archivePrefix = {arXiv},
       eprint = {2502.02642},
 primaryClass = {astro-ph.GA},
       adsurl = {https://ui.adsabs.harvard.edu/abs/2025ApJ...983L..10Z}
}

@ARTICLE{2025MNRAS.542.1331D,
       author = {{Dillamore}, Adam M. and {Sanders}, Jason L.},
        title = "{Bar-driven dispersal of Galactic substructure}",
      journal = {\mnras},
         year = 2025,
        month = sep,
       volume = {542},
       number = {2},
        pages = {1331-1346},
          doi = {10.1093/mnras/staf1264},
archivePrefix = {arXiv},
       eprint = {2506.09117},
 primaryClass = {astro-ph.GA},
       adsurl = {https://ui.adsabs.harvard.edu/abs/2025MNRAS.542.1331D}
}

@ARTICLE{2021MNRAS.500.4710C,
       author = {{Chiba}, Rimpei and {Friske}, Jennifer K.~S. and {Sch{\"o}nrich}, Ralph},
        title = "{Resonance sweeping by a decelerating Galactic bar}",
      journal = {\mnras},
         year = 2021,
        month = jan,
       volume = {500},
       number = {4},
        pages = {4710-4729},
          doi = {10.1093/mnras/staa3585},
archivePrefix = {arXiv},
       eprint = {1912.04304},
 primaryClass = {astro-ph.GA},
       adsurl = {https://ui.adsabs.harvard.edu/abs/2021MNRAS.500.4710C}
}

\begin{appendix}

\section{Acknowledgements}\label{app:acknowledgements}

\begin{acknowledgements}
The authors gratefully acknowledge the anonymous referee for providing the comments that improved our work. TT and DT thank J. Baba for fruitful discussions on Galactic dynamics. This work has been supported by the Tokyo Center For Excellence Project, Tokyo Metropolitan University. TT acknowledges the support by JSPS KAKENHI Grant Nos. 22K18280 and 23H00132. DT acknowledges financial support from JSPS Research Fellowship for Young Scientists and accompanying Grants-in-Aid for JSPS Fellows (23KJ2149). PdL and ARB acknowledge partial funding from the European Union's Horizon 2020 research and innovation program under SPACE-H2020 grant agreement number 101004214 (EXPLORE project). 
\\
This work has made use of data from the European Space Agency (ESA) mission \textit{Gaia} (https://www.cosmos.esa.int/gaia), processed by the \textit{Gaia} Data Processing and Analysis Consortium (DPAC, https://www.cosmos.esa.int/web/gaia/dpac/consortium). Funding for the DPAC has been provided by national institutions, in particular the institutions participating in the Gaia Multilateral Agreement. 
\end{acknowledgements}

\section{Deconvolution of the age histogram}\label{app:deconvolve}

\begin{figure}
\centering 
\includegraphics[width=9cm]{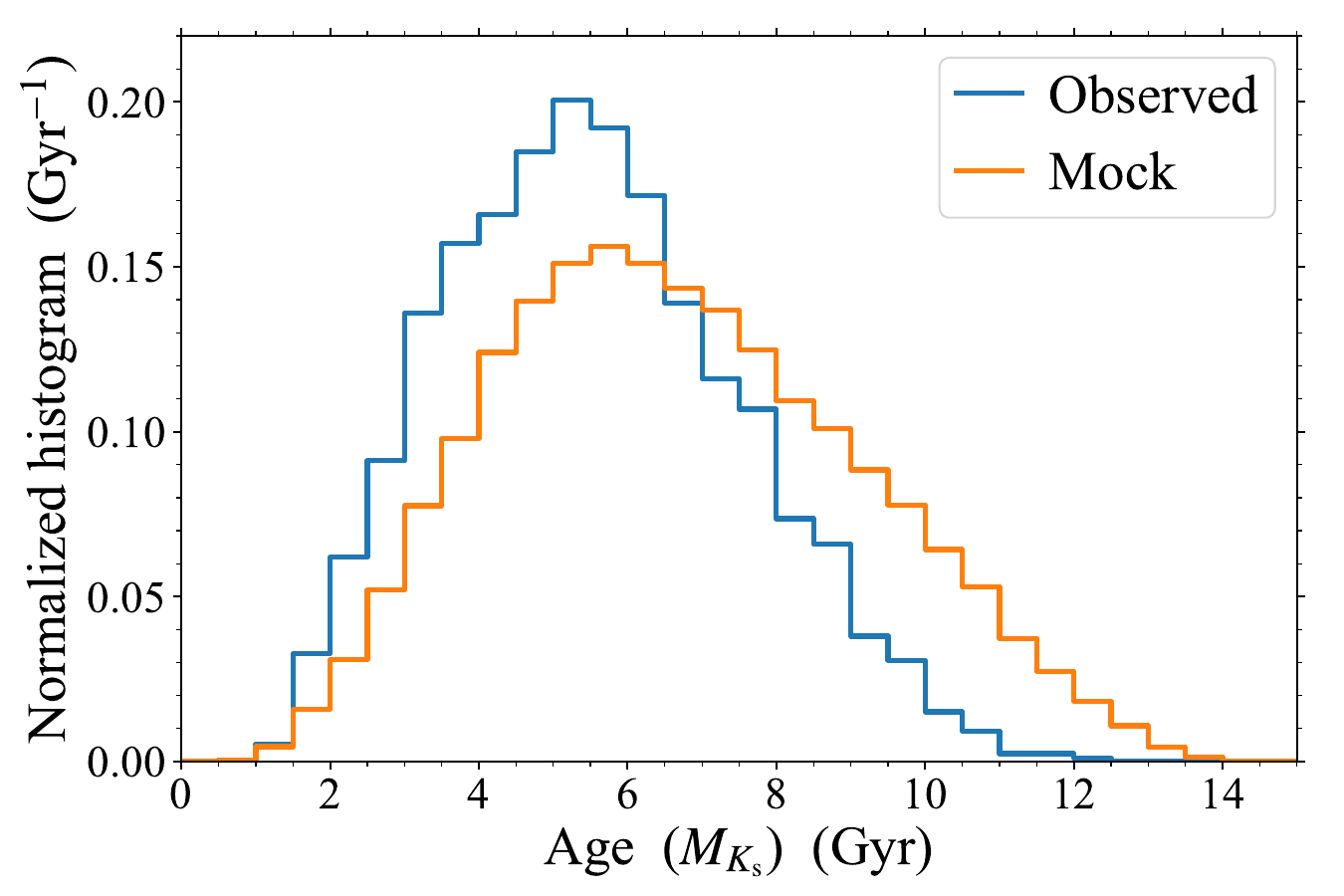}
\caption{Observed histogram of solar twin ages. Histograms shows the ages of stars with the relative age error smaller than $50\%$, determined using $M_{K_{\mathrm{s}}}$ for the observed (blue) and mock (orange) samples. See Fig.~\ref{fig:AgeHist0_logg_G0} for results using $\log g$ and those using $M_{G}$. }
\label{fig:AgeHist0_K0}
\end{figure}

In this Appendix, we provide a detailed description of the deconvolution procedure outlined in Sect.~\ref{sec:DataSet} to derive the ``intrinsic'' age distribution of migrated stars. For this purpose, just showing the histogram of solar twin ages (blue histogram in Fig.~\ref{fig:AgeHist0_K0}) is insufficient, given that the solar twin selection effects alter the histogram~(see Sect.~4.3.2 in \citetalias{Paper1}). Such effects should be corrected for, or at minimum accounted for, when analyzing datasets from larger surveys~\citep{2016ARA&A..54..529B}. 

To characterize the observational selection effects of our solar twin catalog, \citetalias{Paper1} constructed a mock solar twin sample. Stellar ages and initial metallicities were randomly sampled from uniform distributions, and initial masses were sampled from the \citet{2001MNRAS.322..231K} initial mass function. Synthetic stellar parameters and photometric quantities were then generated using the PARSEC isochrones. Simulated observational uncertainties and the same selection criteria as for the real sample were applied, thereby embedding the expected selection effects for the observed solar twins catalog in the mock catalog (see Sect.~4.3.1 in \citetalias{Paper1} for more details). 

The selection effects in our solar twin sample are illustrated in the orange histogram in Fig.~\ref{fig:AgeHist0_K0}, which shows the age histogram of mock solar twins assuming a flat intrinsic age distribution. The figure illustrates, for example, that most solar-type stars older than ${\sim }10\ur{Gyr}$ have already evolved off the main sequence and therefore no longer satisfy the solar twin criteria (i.e., $T_{\mathrm{eff}}$, $\log g$, and [M/H] are around the solar values). Consequently, even though the intrinsic age distribution is flat, the mock solar twin histogram contains very few stars at such old ages. 

\begin{figure}
\centering 
\includegraphics[width=9cm]{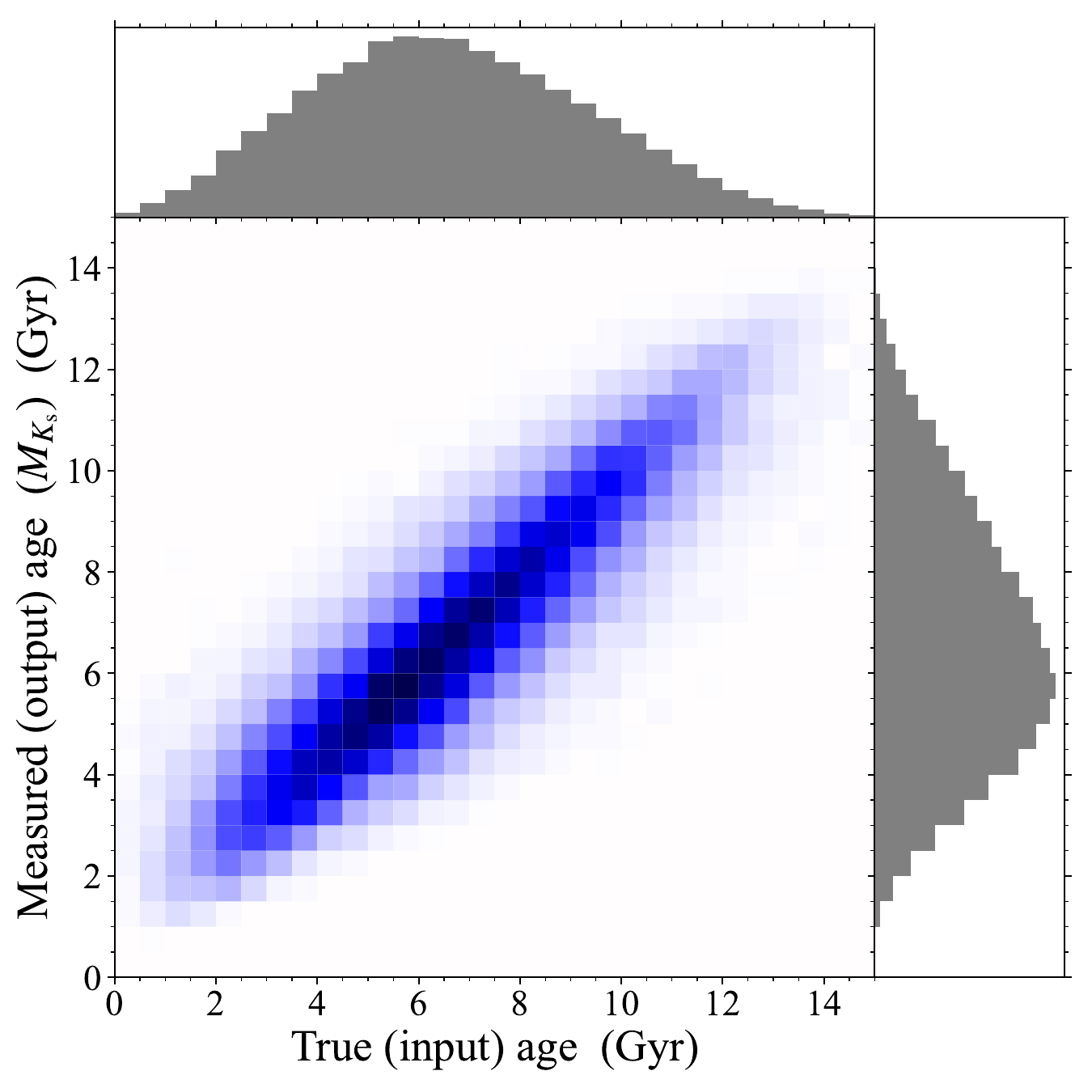}
\caption{Selection function for solar twin ages determined with $M_{K_{\mathrm{s}}}$. 2D histogram in the central panel shows the distribution of the input true and output estimated ages for the mock solar twins. Top and right panels show the normalized marginal histograms for the input and output ages, respectively. See Fig.~\ref{fig:SelecFunc_logg_G0} for results using $\log g$ and those using $M_{G}$. }
\label{fig:SelecFunc_K0}
\end{figure}

Here, we describe how we corrected for the selection effects from the observed solar twin age histogram. The 2D histogram in Fig.~\ref{fig:SelecFunc_K0} shows the selection function matrix $S$ for our solar twin ages determined with $M_{K_{\mathrm{s}}}$. The matrix was constructed by binning the true (input) and inferred (output) ages of $75{,}588$ mock stars generated by \citetalias{Paper1} on a regular grid with $0.5\ur{Gyr}$ bins. We remind that the mock stars were initially drawn from a flat intrinsic age distribution before applying the solar twin selection criteria. By construction, only mock stars that satisfy the solar twin selection criteria are included in this 2D histogram. The matrix $S$ is therefore not normalized by column or row. 

With this definition of matrix $S$, any observed solar twin age histogram $\vec{h}_{\mathrm{obs}}$ is related to the corresponding intrinsic age histogram $\vec{h}_{\mathrm{true}}$ through the selection function matrix $S$ derived from the mock catalog: 
\begin{equation}
\vec{h}_{\mathrm{obs}}=S\vec{h}_{\mathrm{true}}\text{.}
\end{equation}
In the current analysis, we estimate $\vec{h}_{\mathrm{true}}$ for the parent population that gave rise to the observed solar twins, by deconvolving the age distribution $\vec{h}_{\mathrm{obs}}$ of the solar twins actually observed. We employed two independent deconvolution approaches: 
\begin{enumerate}
\item Regularized least-squares (RLS) inversion. In this approach, the residual between the observed histogram $\vec{h}_{\mathrm{obs}}$ and the recovered $S\vec{h}_{\mathrm{true}}$ was minimized under Tikhonov regularization with a discrete difference operator $D$ and a regularization parameter $\lambda $ to enforce the smoothness in $\vec{h}_{\mathrm{true}}$: 
\begin{equation}
\min _{\vec{h}_{\mathrm{true}}\geq 0}\left[\norm{\vec{w}\odot (\vec{h}_{\mathrm{obs}}-S\vec{h}_{\mathrm{true}})}_{2}^{2}+\lambda \norm{D\vec{h}_{\mathrm{true}}}_{2}^{2}\right]\text{,}
\end{equation}
where the weight vector was defined as $\vec{w}=\vec{1}/\sqrt{\vec{h}_{\mathrm{obs}}+\vec{1}}$. The addition of the constant term prevents the weights from diverging in bins with very small $\vec{h}_{\mathrm{obs}}$ values at old ages. 
\item Richardson-Lucy (RL) iterative deconvolution~\citep{1972JOSA...62...55R,1974AJ.....79..745L}. In this approach, $\vec{h}_{\mathrm{obs}}$ was iteratively updated according to
\begin{equation}
\vec{h}_{\mathrm{true}}^{(n+1)} = \vec{h}_{\mathrm{true}}^{(n)} \odot \frac{S^{\mathrm{T}}\left(\vec{h}_{\mathrm{obs}}\oslash (S\vec{h}_{\mathrm{true}}^{(n)})\right)}{S^{\mathrm{T}}\vec{1}}\text{,}
\end{equation}
starting from a flat initial guess $\vec{h}_{\mathrm{true}}^{(0)}$. The total iteration number $N$ serves as a hyperparameter that implicitly regularizes the solution through early stopping. 
\end{enumerate}
Both methods were applied to the observed histogram $\vec{h}_{\mathrm{obs}}$ while varying the corresponding hyperparameters ($\lambda $ for the RLS method and $N$ for the RL method). 

\begin{figure}
\centering 
\includegraphics[width=9cm]{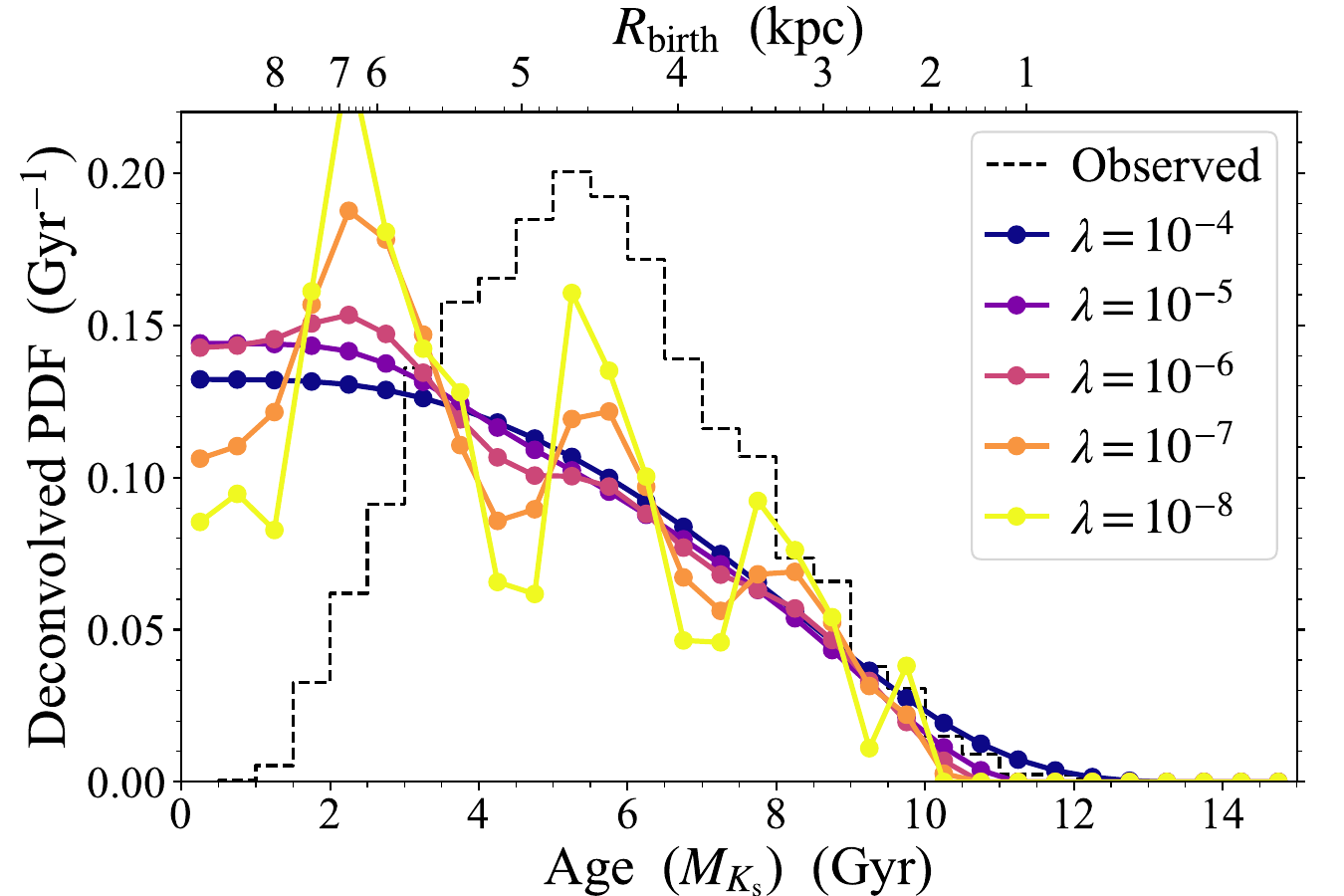}
\caption{Dependence of the deconvolved $M_{K_{\mathrm{s}}}$-based age PDF obtained with the RLS method on the hyperparameter $\lambda $. }
\label{fig:DeconvHist_K0_RLS}
\end{figure}

\begin{figure}
\centering 
\includegraphics[width=9cm]{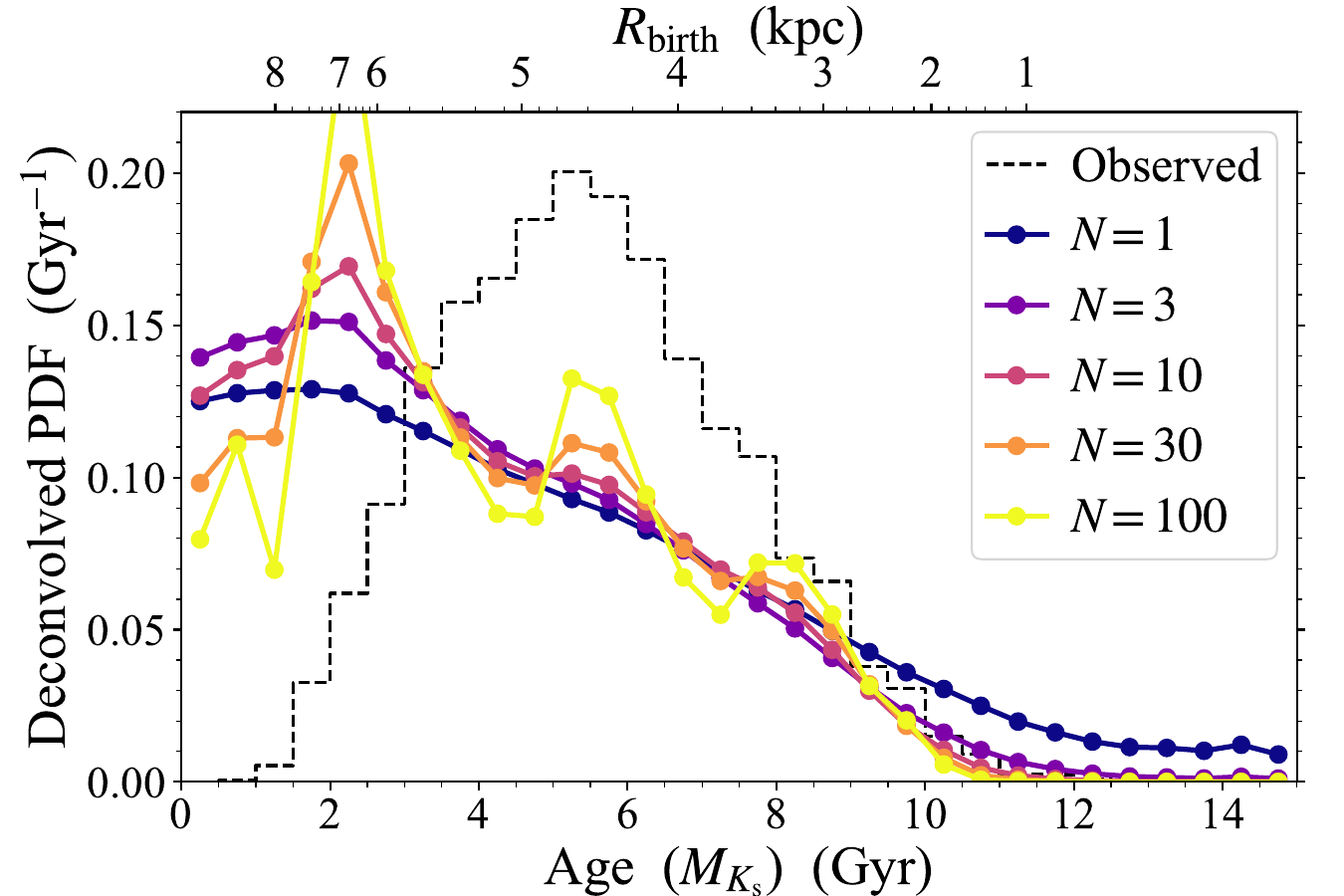}
\caption{Dependence of the deconvolved $M_{K_{\mathrm{s}}}$-based age PDF obtained with the RL method on the hyperparameter $N$. }
\label{fig:DeconvHist_K0_RL}
\end{figure}

To see the behavior of both methods, Figs.~\ref{fig:DeconvHist_K0_RLS} and \ref{fig:DeconvHist_K0_RL} show the deconvolved intrinsic age distributions $\vec{h}_{\mathrm{true}}$ obtained with the RLS and RL methods, respectively, for different hyperparameter values. In the RLS method, $\vec{h}_{\mathrm{true}}$ remains relatively flat when $\lambda $ is large (i.e., more weight on the smoothness), whereas small-scale, possibly artificial, structures emerge as $\lambda $ decreases. A similar trend is also seen with the RL method, where increasing $N$ (i.e., repeating the iterative update) produces a progressively bumpier $\vec{h}_{\mathrm{true}}$. 

\begin{figure}
\centering 
\includegraphics[width=9cm]{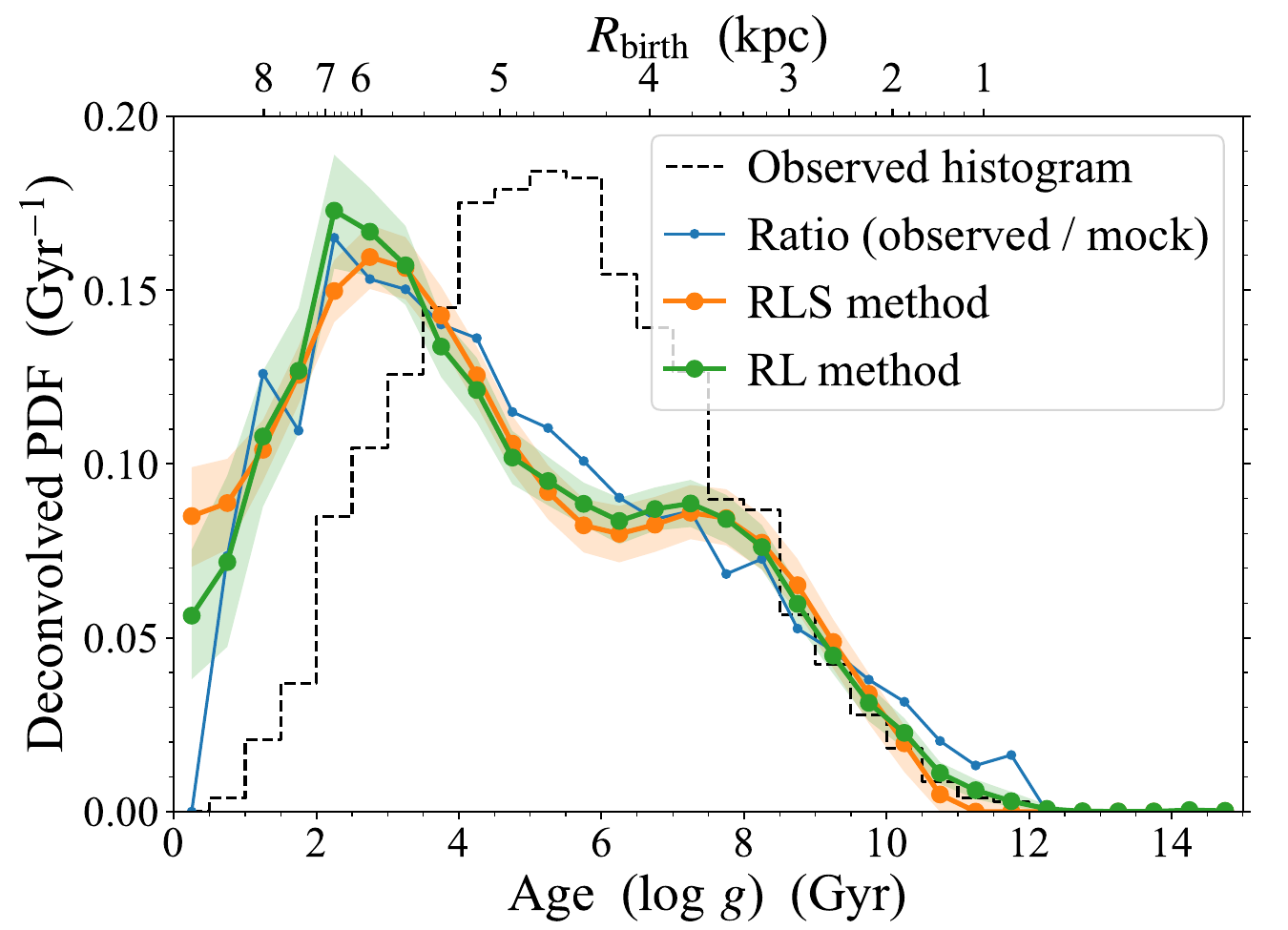}
\includegraphics[width=9cm]{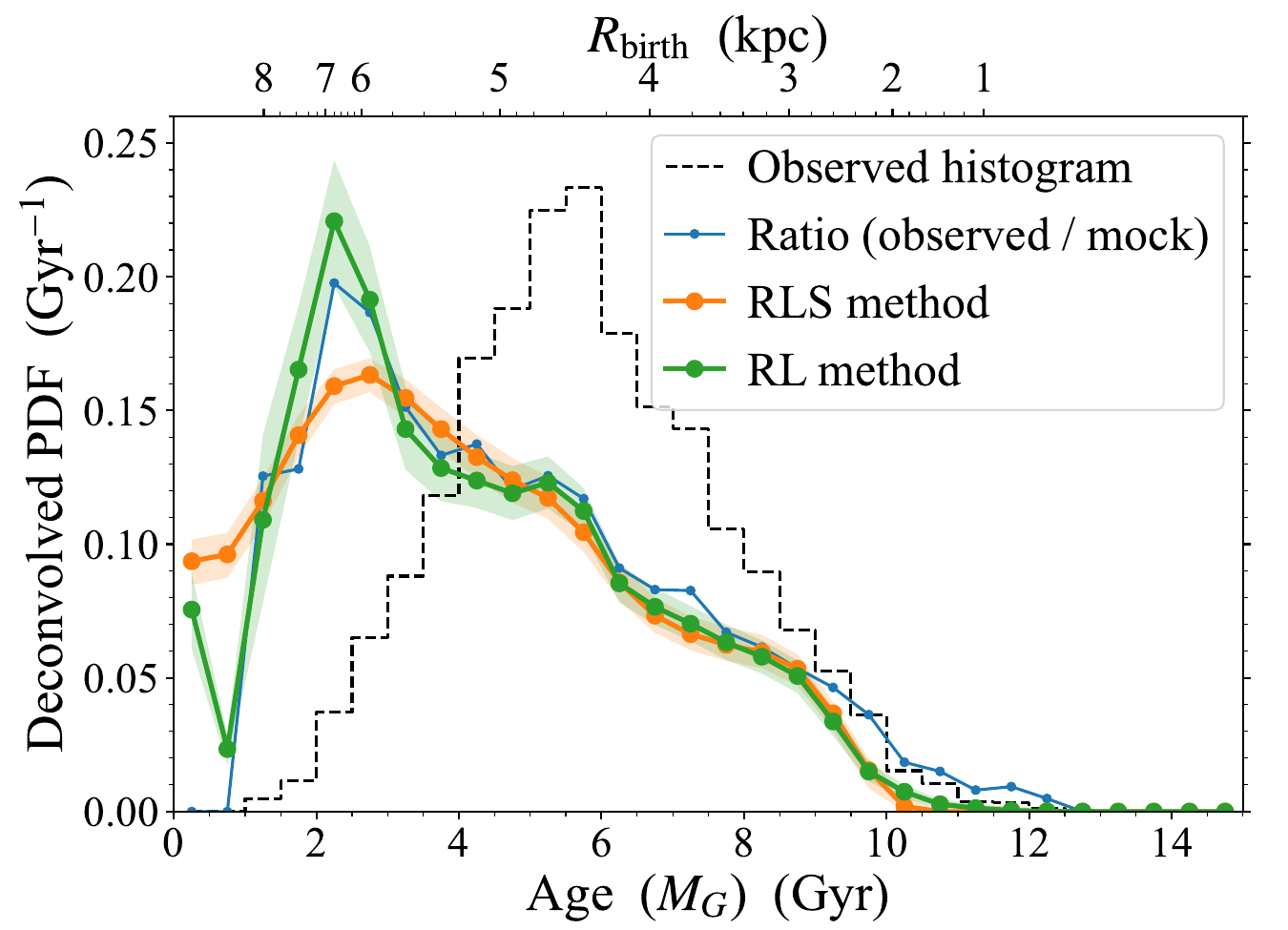}
\caption{The same as Fig.~\ref{fig:DeconvHist_K0} but for the age determined with $\log g$ (top) and $M_{G}$ (bottom). }
\label{fig:DeconvHist_logg_G0}
\end{figure}

To choose the optimal hyperparameter value, we divided $\vec{h}_{\mathrm{obs}}$ into five contiguous blocks and performed leave-one-block-out cross validation. Across a grid of hyperparameter values, we chose the one minimizing the residual $\norm{\vec{w}\odot (\vec{h}_{\mathrm{obs}}-S\vec{h}_{\mathrm{true}})}_{2}^{2}$ in the validation set. For the $M_{K_{\mathrm{s}}}$-based ages, the optimal values were $\lambda =10^{-5.8}$ for the RLS method and $N=16$ for the RL method, which were used to produce Fig.~\ref{fig:DeconvHist_K0}. We chose $\lambda =10^{-6.5}$ and $N=20$ for the $\log g$-based ages, and $\lambda =10^{-6.1}$ and $N=20$ for the $M_{G}$-based ages in Fig.~\ref{fig:DeconvHist_logg_G0}.

\section{Impact of finite-sample statistics}\label{app:StatBias}

\begin{figure}
\centering 
\includegraphics[width=9cm]{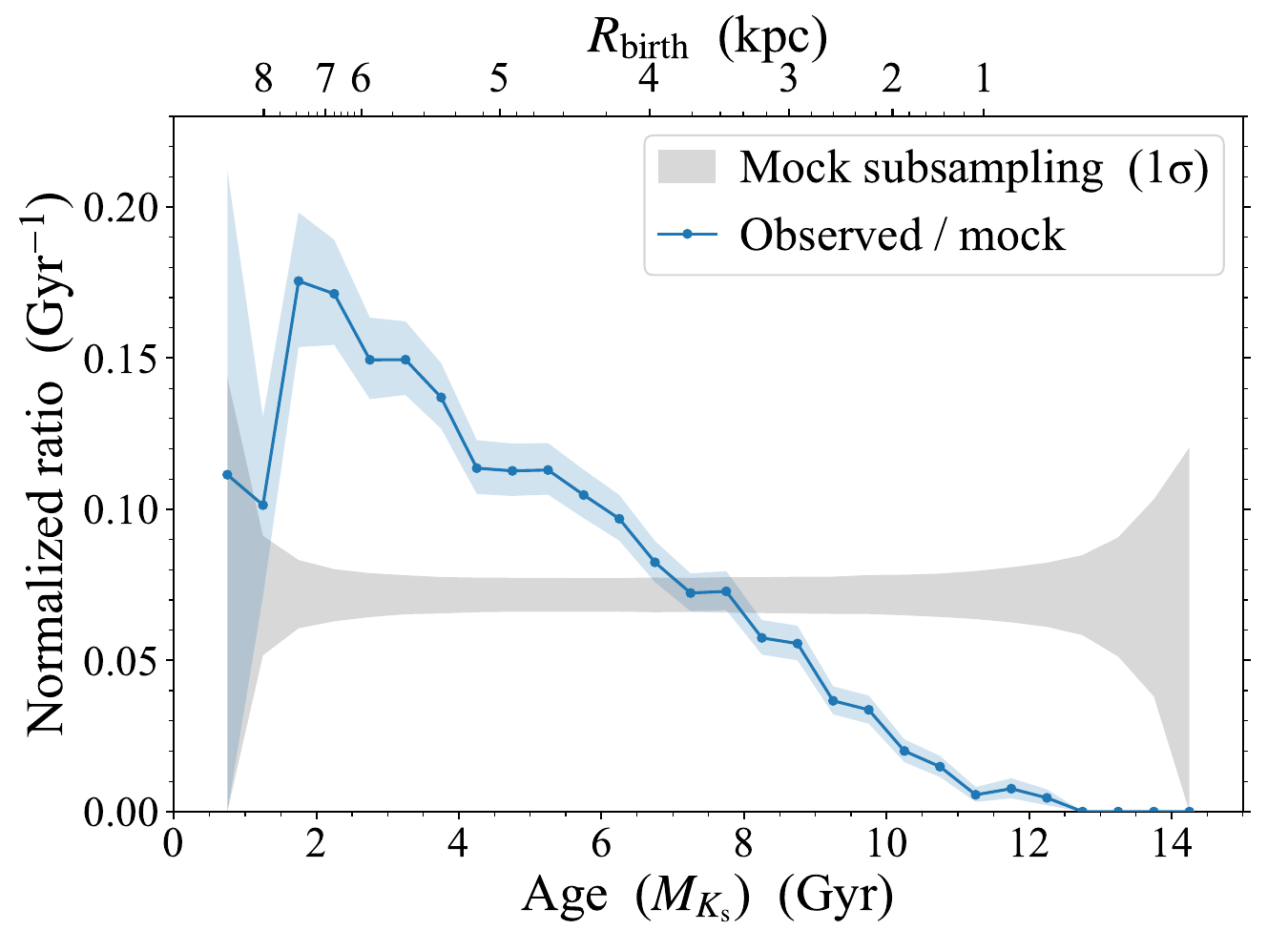}
\caption{Finite-sample effects on the observed-to-mock age histogram ratio. Blue line shows the normalized ratio between the observed and mock age histograms for solar twins based on $M_{K_{\mathrm{s}}}$, identical to that shown in Fig.~\ref{fig:DeconvHist_K0}. Blue shaded region indicates the statistical uncertainty on this ratio, estimated from $10^{5}$ Monte Carlo realizations assuming Poisson statistics for the observed age histogram. Gray shaded region represents the $1\sigma $ sampling variance expected from finite-sample statistics, estimated by repeatedly drawing random subsamples from the mock catalog with the same number of stars as in the observed sample and computing the corresponding subsample-to-mock ratios. }
\label{fig:StatBias_K0}
\end{figure}

Our observed solar twin sample consists of a finite number of stars ($6{,}594$ stars), and therefore any age distribution inferred from this sample is subject to statistical fluctuations. In the following, we use the normalized ratio between the observed and mock age histograms (blue line in Fig.~\ref{fig:DeconvHist_K0}) as a rough estimate for the intrinsic age distribution to examine the impact of finite-sample statistics. In the main text, statistical uncertainties in the deconvolved age PDFs (orange and green shaded regions in Fig.~\ref{fig:DeconvHist_K0}) are estimated using Monte Carlo (MC) realizations that assume Poisson statistics for the observed age histogram. In this Appendix, we validate whether this uncertainty estimate is consistent with the sampling variance expected from a finite-sized sample. 

We performed an independent MC subsampling experiment using only the mock catalog, which contains approximately ten times more stars ($75{,}588$ stars) than the observed sample. From the full mock catalog, we repeatedly chose random subsamples with the same number of stars as in the observed sample and computed the normalized ratio between each subsample and the full mock catalog, in the same way as for the observed data. After $10^{4}$ realizations, the distribution of these ratios quantifies the sampling variance expected for the observed sample. 

Gray shaded region in Fig.~\ref{fig:StatBias_K0} shows the $1\sigma $ sampling variance derived from the mock subsampling. The typical fractional $1\sigma $ width is $9\%$ (gray region), which is comparable to the typical statistical uncertainty of $10\%$ estimated from the MC realizations of the observed sample (blue region). This agreement indicates that our estimation of the statistical uncertainty in the deconvolved intrinsic age distribution provides a reasonable approximation to the sampling variance arising from finite-sample statistics. However, it should be noted that the presence of the bump around $4\text{--}6\ur{Gyr}$ ago may become marginal when these statistical uncertainties are taken into account, especially given that the RLS method does not detect such a bump feature. Nonetheless, we emphasize that Solar twins with ages comparable to that of the Sun do not constitute a negligible population in the solar neighborhood.

\section{Additional figures}\label{app:figures}

\begin{figure}
\centering 
\includegraphics[width=9cm]{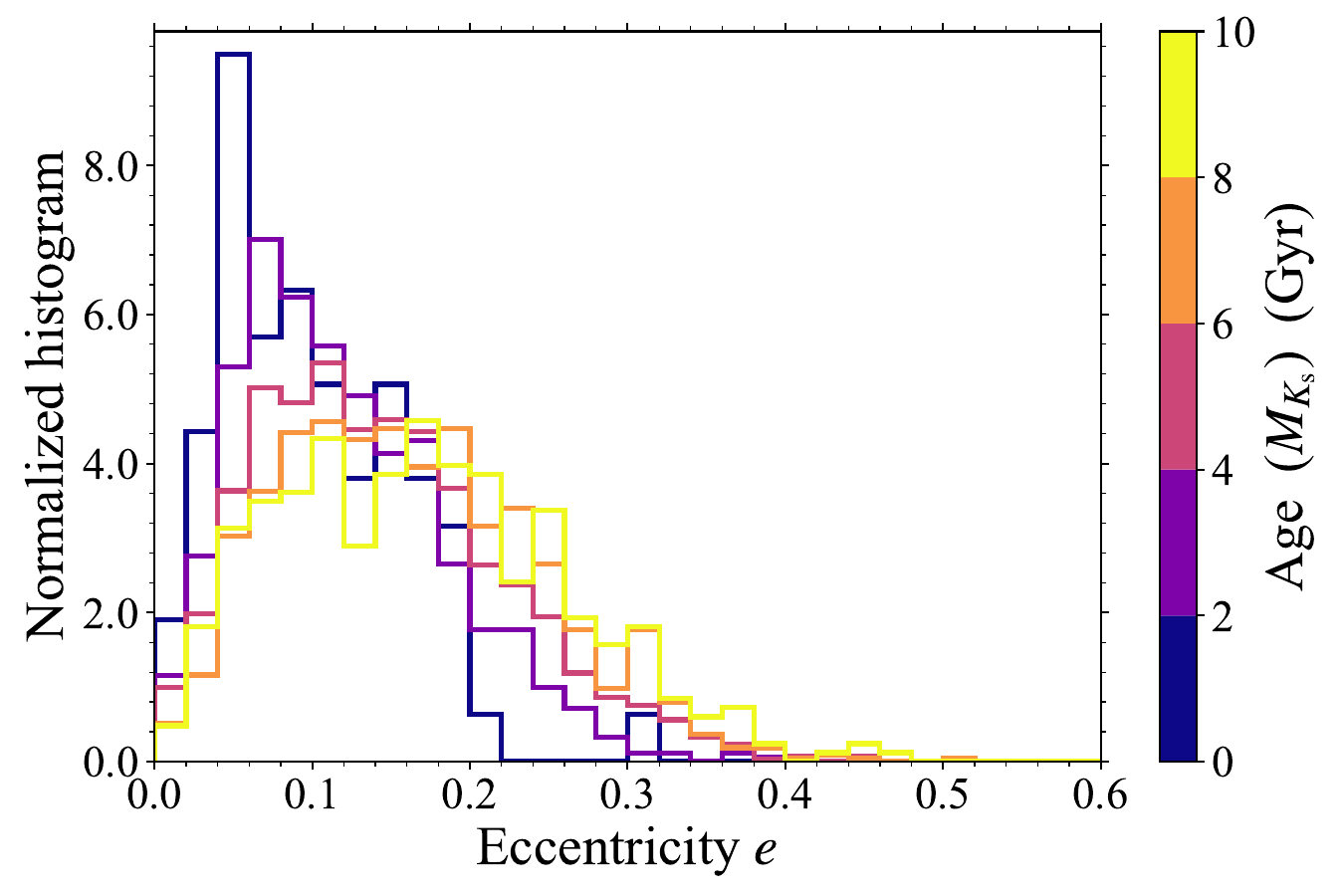}
\includegraphics[width=9cm]{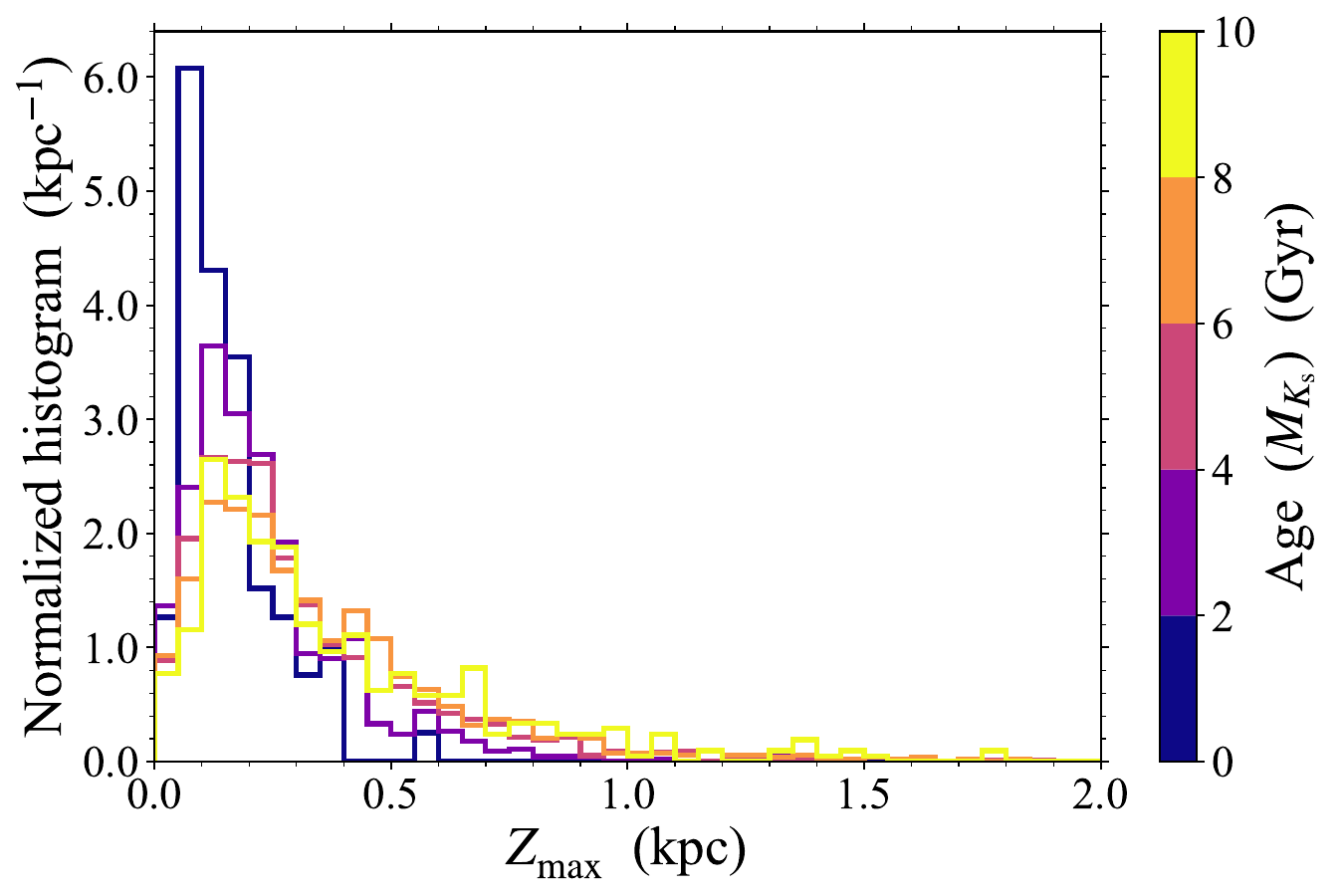}
\caption{The same as Fig.~\ref{fig:OrbitParams_K0_Rg} but for the eccentricity $e$ (left) and the maximum vertical distance from the Galactic plane $Z_{\mathrm{max}}$ (right). }
\label{fig:OrbitParams_K0_ecc_zmax}
\end{figure}

\begin{figure}
\centering 
\includegraphics[width=9cm]{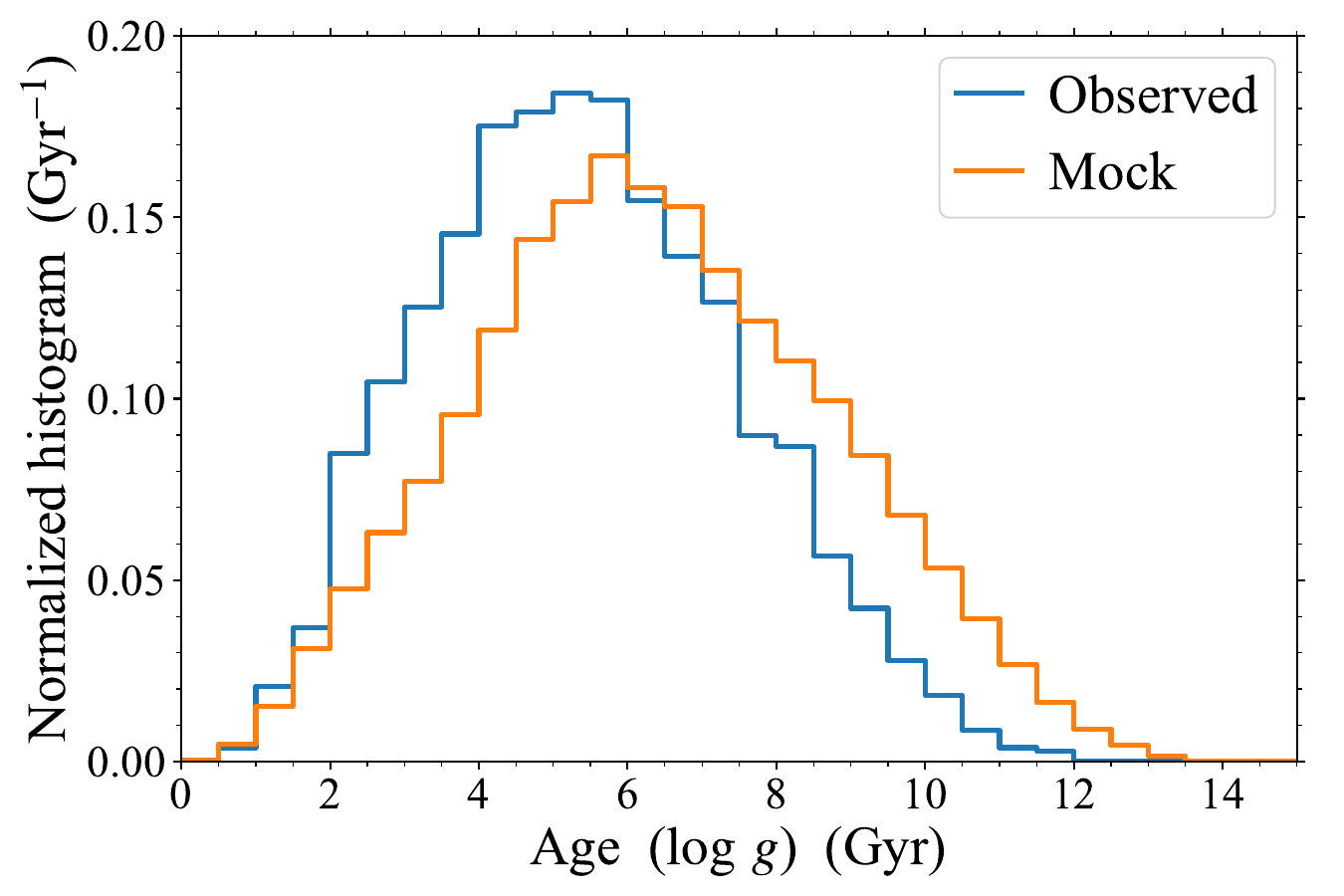}
\includegraphics[width=9cm]{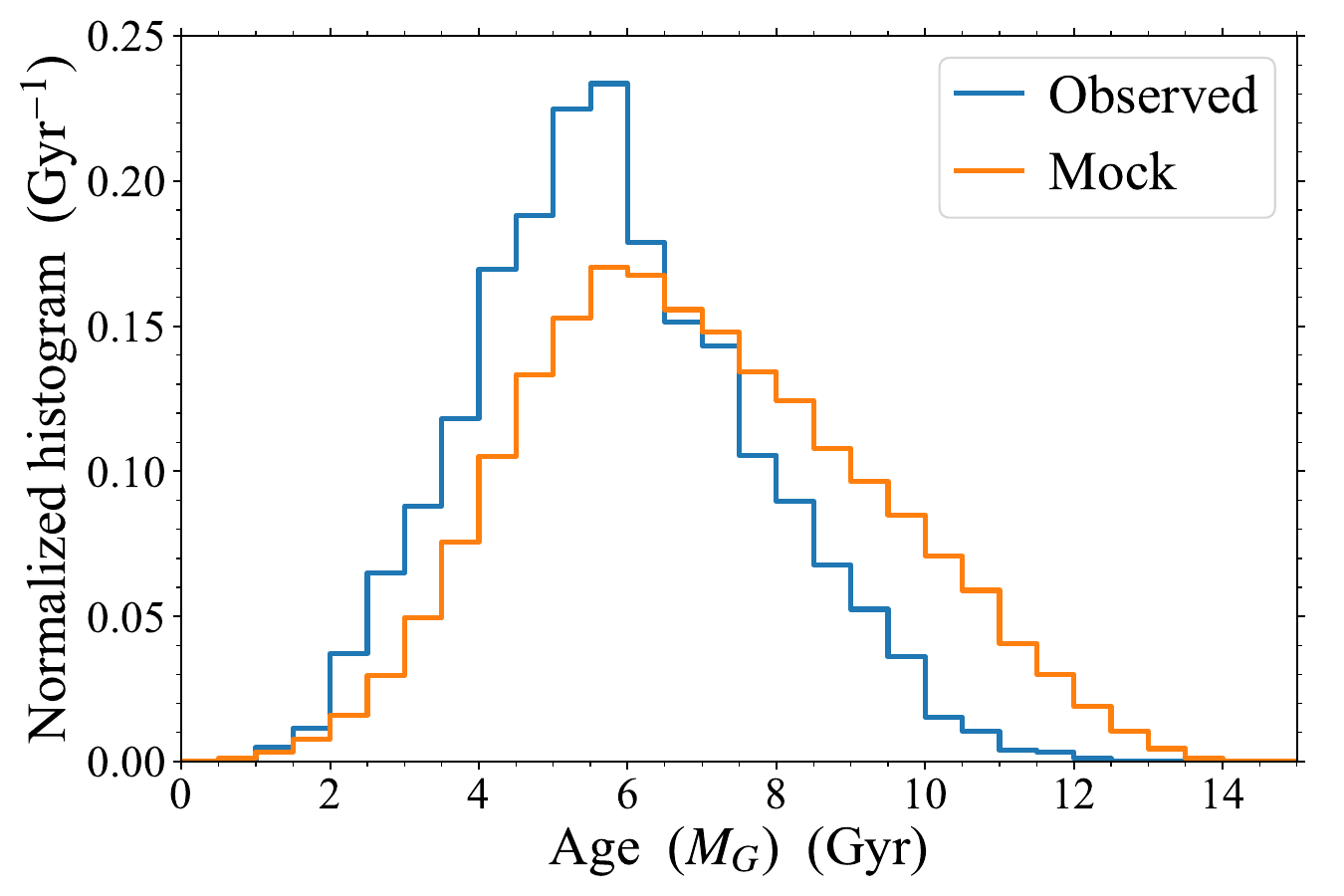}
\caption{The same as Fig.~\ref{fig:AgeHist0_K0} but for the age determined with $\log g$ (top) and $M_{G}$ (bottom). }
\label{fig:AgeHist0_logg_G0}
\end{figure}

\begin{figure}
\centering 
\includegraphics[width=9cm]{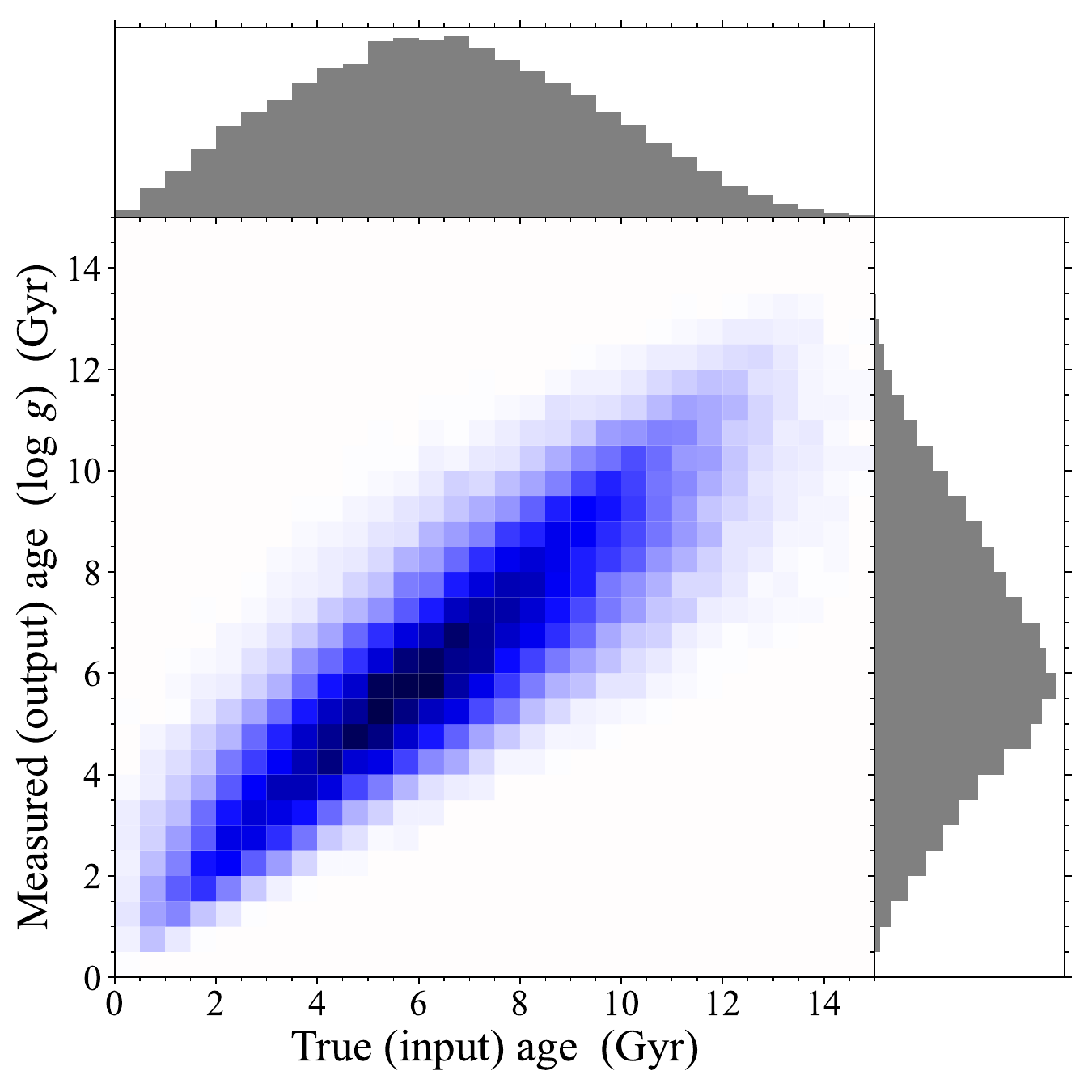}
\includegraphics[width=9cm]{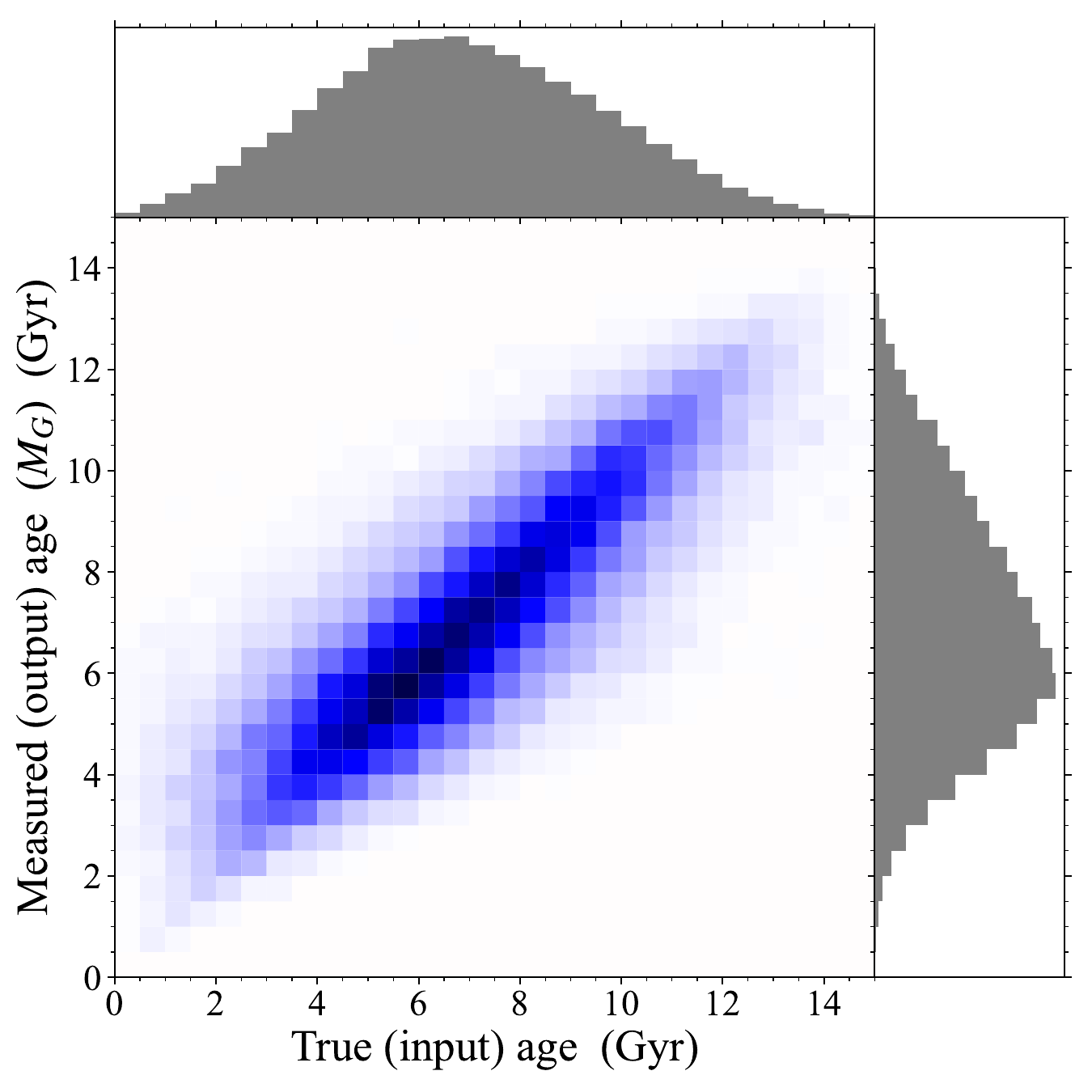}
\caption{The same as Fig.~\ref{fig:SelecFunc_K0} but for the age determined with $\log g$ (top) and $M_{G}$ (bottom). }
\label{fig:SelecFunc_logg_G0}
\end{figure}

\end{appendix}

\end{document}